\documentclass[aps,pra,notitlepage,singlecolumn,superscriptaddress,nofootinbib,
amsmath,amssymb,
]{revtex4-1}

\usepackage{epsfig}
\usepackage{soul} % use colored text in latex
\usepackage{graphicx} % enhanced support for graphics
\usepackage[table]{xcolor}% http://ctan.org/pkg/xcolor
\usepackage{diagbox}

\usepackage{amsthm}
\usepackage{amsmath}
\usepackage{amssymb}
\usepackage{bbm,amsfonts}
\usepackage{nomencl}
\usepackage{amstext} % Lots of math symbols and environments
\usepackage{setspace}
\usepackage{mathtools}

\usepackage[version=4]{mhchem} % for chemistry

\usepackage{url}            % simple URL typesetting
\usepackage{booktabs}       % professional-quality tables
\usepackage{nicefrac}       % compact symbols for 1/2, etc.
\usepackage{microtype}      % microtypography
\usepackage{lipsum}
\usepackage{dsfont} % For getting the cool identity operator

\usepackage{comment}

\usepackage{algorithm}
\usepackage[noend]{algpseudocode}

\usepackage{hyperref}
\hypersetup{colorlinks=true,citecolor={blue}}

\usepackage{dcolumn}% Align table columns on decimal point
\usepackage{bm}% bold math

\usepackage{multirow}

\ExplSyntaxOn

\keys_define:nn { miguel/label }
 {
  label   .tl_set:N = \l_miguel_label_tl,
  unknown .code:n   = \clist_put_right:Nx \l_miguel_label_clist
                       { \l_keys_key_tl = \exp_not:n { #1 } }
 }
\clist_new:N \l_miguel_label_clist
\box_new:N \l_miguel_label_box
\box_new:N \l_miguel_label_image_box

\NewDocumentCommand{\xincludegraphics}{O{}m}
 {
  \tl_clear:N \l_miguel_label_tl
  \clist_clear:N \l_miguel_label_clist
  \keys_set:nn { miguel/label } { #1 }
  \tl_if_empty:NTF \l_miguel_label_tl
   {
    \miguel_includegraphics:Vn \l_miguel_label_clist { #2 }
   }
   {
    \hbox_set:Nn \l_miguel_label_image_box
     {
      \miguel_includegraphics:Vn \l_miguel_label_clist { #2 }
     }
    \hbox_set:Nn \l_miguel_label_box
     {
      \skip_horizontal:n { -9pt }
      \fcolorbox{white}{white}{\footnotesize \tl_use:N \l_miguel_label_tl}
     }
    \leavevmode
    \box_use:N \l_miguel_label_image_box
    \skip_horizontal:n { -\box_wd:N \l_miguel_label_image_box }
    \hbox_overlap_right:n
     {
      \box_move_up:nn
       {
        \box_ht:N \l_miguel_label_image_box - 
        \box_ht:N \l_miguel_label_box/5
       }
       { \box_use_drop:N \l_miguel_label_box }
     }
    \skip_horizontal:n { \box_wd:N \l_miguel_label_image_box }
   }
 }

\cs_new_protected:Nn \miguel_includegraphics:nn
 {
  \includegraphics[#1]{#2}
 }
\cs_generate_variant:Nn \miguel_includegraphics:nn { V }

\ExplSyntaxOff

\newcommand{\covers}{\triangleright}

\newcommand{\benchname}{\text{CSHOREBench}}
\newcommand{\benchabbr}{\text{Common States and Hamiltonians for ObseRvable Estimation}}

\DeclareMathOperator{\Tr}{Tr}

\def\01{\{0,1\}}

\newcommand{\Exp}{\mathbb{E}}
\newcommand{\Var}{\mathrm{Var}}

\definecolor{applegreen}{rgb}{0.0, 0.5, 0.0}

\global\long\def\argmin{\operatornamewithlimits{argmin}}

\begin{document}
\title{Practical Benchmarking of Randomized Measurement Methods for \\ Quantum Chemistry Hamiltonians}
% Alternate title: Practical benchmarking hybrid quantum-classical routine of estimating quantum Hamiltonians
\author{Arkopal Dutt}
\email{arkopal@mit.edu}
% \affiliation{MIT-IBM Watson AI Lab, Cambridge, MA 02141, USA}
\affiliation{IBM Quantum, IBM Research Cambridge, Cambridge, MA 02141, USA}
\affiliation{Department of Physics, Co-Design Center for Quantum Advantage, Massachusetts Institute of Technology, Cambridge, Massachusetts 02139, USA}

\author{William Kirby}
\affiliation{IBM Quantum, IBM Research Cambridge, Cambridge, MA 02141, USA}

\author{Rudy Raymond}
% \affiliation{IBM Quantum, Thomas J Watson Research Center, Yorktown Heights, New York 10598, USA}
\affiliation{Department of Computer Science, The University of Tokyo, Tokyo 113-8654, Japan}

\author{Charles Hadfield}
\thanks{Former affiliation}
\affiliation{IBM Quantum, Thomas J Watson Research Center, Yorktown Heights, New York 10598, USA}

\author{Sarah Sheldon}
\affiliation{IBM Quantum, Thomas J Watson Research Center, Yorktown Heights, New York 10598, USA}

\author{Isaac L. Chuang}
\affiliation{Department of Physics, Co-Design Center for Quantum Advantage, Massachusetts Institute of Technology, Cambridge, Massachusetts 02139, USA} 
\affiliation{Department of Electrical Engineering and Computer Science, Massachusetts Institute of Technology, Cambridge, Massachusetts 02139, USA}

\author{Antonio Mezzacapo}
\affiliation{IBM Quantum, Thomas J Watson Research Center, Yorktown Heights, New York 10598, USA}

\begin{abstract}
    Many hybrid quantum-classical algorithms for the application of ground state energy estimation in quantum chemistry involve estimating the expectation value of a molecular Hamiltonian with respect to a quantum state through measurements on a quantum device. To guide the selection of measurement methods designed for this observable estimation problem, we propose a benchmark called CSHOREBench (Common States and Hamiltonians for ObseRvable Estimation Benchmark) that assesses the performance of these methods against a set of common molecular Hamiltonians and common states encountered during the runtime of hybrid quantum-classical algorithms. In CSHOREBench, we account for resource utilization of a quantum computer through measurements of a prepared state, and a classical computer through computational runtime spent in proposing measurements and classical post-processing of acquired measurement outcomes. We apply CSHOREBench considering a variety of measurement methods on Hamiltonians of size up to 16 qubits. Our discussion is aided by using the framework of decision diagrams which provides an efficient data structure for various randomized methods and illustrate how to derandomize distributions on decision diagrams. In numerical simulations, we find that the methods of decision diagrams and derandomization are the most preferable. In experiments on IBM quantum devices against small molecules, we observe that decision diagrams reduces the number of measurements made by classical shadows by more than $80\%$, that made by locally biased classical shadows by around $57\%$, and consistently require fewer quantum measurements along with lower classical computational runtime than derandomization. Furthermore, CSHOREBench is empirically efficient to run when considering states of random quantum ansatz with fixed depth.
\end{abstract}

\maketitle

\section{Introduction}\label{sec:intro}
The electronic structure problem, and more specifically the problem of approximating ground states, is one of the outstanding challenges in computational chemistry.
Over nearly the past century, an enormous amount of scholarship has gone into developing classical methods for this task (Hartree-Fock~\cite{purplegiant}, MP2~\cite{purplegiant}, configuration interaction~\cite{purplegiant}, coupled cluster~\cite{purplegiant}, density functional theory~\cite{hohenbergkohn1964,kohnsham1965}, quantum Monte Carlo~\cite{hammond1994qmc}, and DMRG~\cite{chan2002dmrg}, to name some notable examples), and in recent decades a large proportion of scientific HPC resources have been dedicated to solving it (see for example~\cite{nerscworkload}).
The computational intensiveness of the electronic structure problem has contributed to the motivation for developing benchmarks for ranking algorithms across many scientific areas including computational chemistry, including for example QM7 \cite{blum2009qmdata,rupp2012fast}, QM9 \cite{ruddigkeit2012qm9data,ramakrishnan2014quantum}, and W4-11 \cite{karton2011w4}. Several of these listed benchmarks in computational chemistry involve datasets for tasks pertaining to molecular property prediction, but there are also benchmarks that focus on algorithmic or methodological aspects without relying on specific datasets, for example, basis set benchmarking \cite{karton2007basis,marshall2011basisset} for describing electronic structure of molecules, conformal benchmarking \cite{friedrich2017conform} to assess algorithms for exploring the low-energy conformational space of molecules, and reaction path benchmarking \cite{maeda2019reaction} to compare optimization methods for finding chemical reaction pathways.

% Common computation and common objects
The classical computational chemistry benchmarks mentioned above share the idea of testing algorithms against a common computation or prediction task on a set of common objects. This idea is prevalent throughout benchmarking for not only comparing different algorithms but also devices with respect to certain measures of performance. For example, the LINPACK benchmark \cite{dongarra1979linpack,dongarra1987linpack}, which was used to rank the top supercomputers in the world, involves the common task of solving linear systems of equations $Ax=b$ where the input matrix $A$ is a pseudo-random dense matrix. The common computation is solving linear systems of equations in the case of LINPACK and molecular prediction tasks for the computational chemistry benchmarks. Additionally, as part of the benchmark, the performance of different algorithms is assessed on the chosen task by testing it on a set of objects. In the case of LINPACK, these objects are pseudo-random dense matrices. In the QM databases, the objects are small organic molecules  \cite{blum2009qmdata,rupp2012fast,ruddigkeit2012qm9data}. The hardness and generality of the set of objects in the benchmark determine how well it will predict the performance of algorithms in practice.
This has been particularly successful in the context of machine learning for image classification (e.g., MNIST \cite{deng2012mnist}, CIFAR \cite{krizhevsky2009cifar}, ImageNet \cite{krizhevsky2012imagenet}) and object detection (e.g., MS COCO \cite{lin2014coco}). In MS COCO, for example, algorithms for object recognition are assessed in the broader context of scene understanding and are tested against images of complex everyday scenes containing common objects in their natural context. By including typically occurring objects in practice as part of the testing suite for benchmarking, there has been an improvement in the development of state-of-art algorithms for object recognition \cite{khan2022transformers,minaee2022image}.

% Potential practical quantum advantage in solving quantum chemistry problems
Returning our attention to the electronic structure problem, on a quantum computer the primary challenge in the classical methods, that of representing highly entangled and correlated wavefunctions, is removed in principle. This, together with the classical hardness of the ground state problem and the enormous amount of resources dedicated to it, has motivated the development of quantum algorithms for the task~\cite{mcardle2020quantum,lee2023evaluating}, although new challenges arise. Numerous methods have been designed, including near-term quantum algorithms such as variational quantum eigensolvers (VQE)~\cite{peruzzo2014vqe,mcclean2016theory,kandala2017hardware,grimsley2019adaptive}, quantum approximate optimization algorithm (QAOA)~\cite{farhi2014quantum,moll2018qaoa,farhi2022quantum}, and quantum subspace expansion (QSE) methods~\cite{mcclean2017subspace,colless2018computation,parrish2019filterdiagonalization,motta2020determining}. Fault-tolerant algorithms for ground state estimation, which are aimed at future high-accuracy quantum computers, include quantum phase estimation (QPE) \cite{kitaev1995phaseestimation}, its variants \cite{abrams1999eigen,poulin2009ground}, algorithm in \cite{ge2019fast} which uses linear combination of unitaries (LCU) \cite{childs2012lcu}, and those using quantum signal processing \cite{gilyen2019qsvt,lin2020near,lin2022ground,ding2023even}.
On currently existing and near-term quantum computers, without error correction, near-term algorithms are preferred for use instead of fault-tolerant algorithms whose circuit depths and qubit counts will require error correction. These near-term algorithms including VQE are typically hybrid quantum-classical algorithms involving sequential rounds of measurements of parametrized quantum circuits or short time Hamiltonian simulation and classical post-processing along with classical optimization. 

A common subroutine across many of these algorithms is that of observable estimation or estimating $\Tr(\rho H)$ (e.g.,~\cite{du2010nmr,lanyon2010towards,wang2015quantum,omalley2016scalable,shen2017quantum,paesani2017bayesian,hempel2018trappedion,santagati2018witnessing,colless2018spectra,dumitrescu2018atomicnucleus,kokail2019selfverifying,kandala2019errormitigation,ganzhorn2019gate,sagastizabal2019experimental,mccaskey2019quantum,smart2019quantum,nam2020trappedion,arute2020hartreefock,kreshchuk2021blfqshort,lotstedt2021calculation,kiss2022quantum}) for a given $n$-qubit quantum state $\rho$ resulting from a short depth quantum circuit and an $n$-qubit Hamiltonian $H$ (or in general any observable). Physical Hamiltonians $H$ can be decomposed into a linear combination of $L$ $n$-qubit Pauli operators: they form a basis for the Hermitian operators, and local observables have polynomial-sized decompositions in the Pauli basis. Other decompositions of $H$ include LCU~\cite{childs2012lcu,kirby2022fermiontoqubit} or one-sparse matrices~\cite{aharonov2003adiabatic,berry2007sparse,childs2011stardecompositions}) but these are impractical in the near-term because of the relatively complex quantum circuits required to estimate expectation values of the terms. In contrast for Pauli decompositions, we could estimate $\Tr(\rho H)$ simply by estimating $\Tr(\rho Q)$ independently~\cite{peruzzo2014vqe} for each of the Pauli terms $Q$ in the Pauli decomposition of $H$.

However, this procedure is typically inefficient as (in general) subsets of Pauli terms will commute and thus be co-measurable. Generally, however, a commuting set of Paulis can only be simultaneously measured by applying a depth-$\Theta(n)$ Clifford circuit to map them to their common eigenbasis. In the near-term, when circuit depth is at a premium due to the lack of error correction, it is desirable to use all of the circuit depth for the state preparation instead of measurements. \emph{Locally} commuting Pauli operators, i.e., operators that have all non-identity single-qubit Pauli matrices in common are used instead. They can be measured simultaneously in the same local basis by applying one layer of single-qubit gates followed by measurement in the computational basis and these are the type of measurements we consider access to in this paper.
% \arko{I am changing circuit volume to circuit depth in the above sentences but please check.}

In recent years, two main approaches for the observable estimation problem using local Pauli measurements have emerged: (i) randomized measurements~\cite{huang2020predicting,hadfield2020measurements,hillmich2021decision,lukens2021bayesian,koh2022classical,elben2023randomized} in which local Pauli measurement bases are drawn from a distribution over the $n$-qubit Pauli operators~\cite{huang2020predicting,hadfield2020measurements,hillmich2021decision} or generated via a sampling procedure that does not require explicit access to the distribution~\cite{hadfield2021adaptive}, and (ii) grouping methods~\cite{gokhale2020commuting,yen2020compatible,verteletskyi2020measurement,crawford2021efficient,wu2023overlapped,yen2023deterministic,shlosberg2023adaptiveestimation}, which combine Pauli terms into locally compatible subsets for simultaneous measurement either systematically \cite{crawford2021efficient,wu2023overlapped} or using ad-hoc heuristics~\cite{kandala2017hardware, hempel2018trappedion,verteletskyi2020measurement}. There also other approaches: \cite{acharya2021informationally} uses a set of informationally complete positive operator-valued measurements to solve the observable estimation problem and \cite{huang2021efficient} obtains deterministic sequences of Pauli measurements to be made by derandomizing randomized measurements. Notably, among the listed methods are those based on classical shadows \cite{huang2020predicting,hadfield2020measurements,hillmich2021decision,wu2023overlapped} which are asymptotically optimal \cite{huang2020predicting} requiring only $O(3^w \log L)$ measurements for a Hamiltonian with $L$ Paulis in its Pauli decomposition and maximum number of non-identity Paulis in any Pauli term being $w$. Despite the potential of these methods in ideal scenarios, little is known about their behavior on quantum devices in presence of noise. Experimental studies have only been carried out so far on small molecular Hamiltonians \cite{struchalin2021experimental} or quantum states over few qubits \cite{zhang2021experimental}. 

Given the large suite of options, a natural question at this point is: How do we systematically select measurement methods for the common quantum computation of estimating $\Tr(\rho H)$ in hybrid quantum-classical algorithms? One way to tackle this is to follow the classical approach and propose a benchmark. This is not without precedent on the quantum side: recently a ``quantum LINPACK'' benchmark \cite{dong2021quantumlinpack} was proposed for ranking computational power of quantum computers; in direct analogy to LINPACK, it involves solving the quantum linear system problem. A challenge in designing benchmarks is to have predictive power regarding the performance of the candidate algorithms beyond the dataset tested against. % and especially in the case of quantum computers where it is still unclear what they would be economically useful for.

In this work, we take a similar approach to classical benchmarks by considering the common quantum computation task of observable estimation $\Tr(\rho H)$ on a set of common chemistry Hamiltonians and quantum states. In analogy to classical computational chemistry benchmarks~\cite{blum2009qmdata,ruddigkeit2012qm9data}, we consider sets of quantum states particular to the problem of ground state estimation as well as those states that naturally occur during the runtime of a hybrid quantum-classical algorithm. This culminates in the proposed benchmark in this work called \benchname:  \benchabbr. 

In addition to commenting on the objects considered as part of the data set of \benchname,~performance metrics used to rank measurement methods on these objects need to be defined. An important aspect in designing or selecting the measurement method and estimator is the amount of resources required. Considering the performance metric of accuracy, one selection criterion is to minimize the number of measurements required on the quantum device in achieving a given accuracy. However, this only takes into account the quantum resources used and may come at a prohibitive computational cost on classical computers in setting up the measurement methods or running the estimator on the data acquired from the experiment step. To capture a representative performance metric for all of the costs associated to a measurement procedure, classical and quantum, we propose a heuristic that incorporates different resources' utilization in the observable estimation problem.

\begin{figure}[ht!]
\centering
\includegraphics[width=0.95\textwidth]{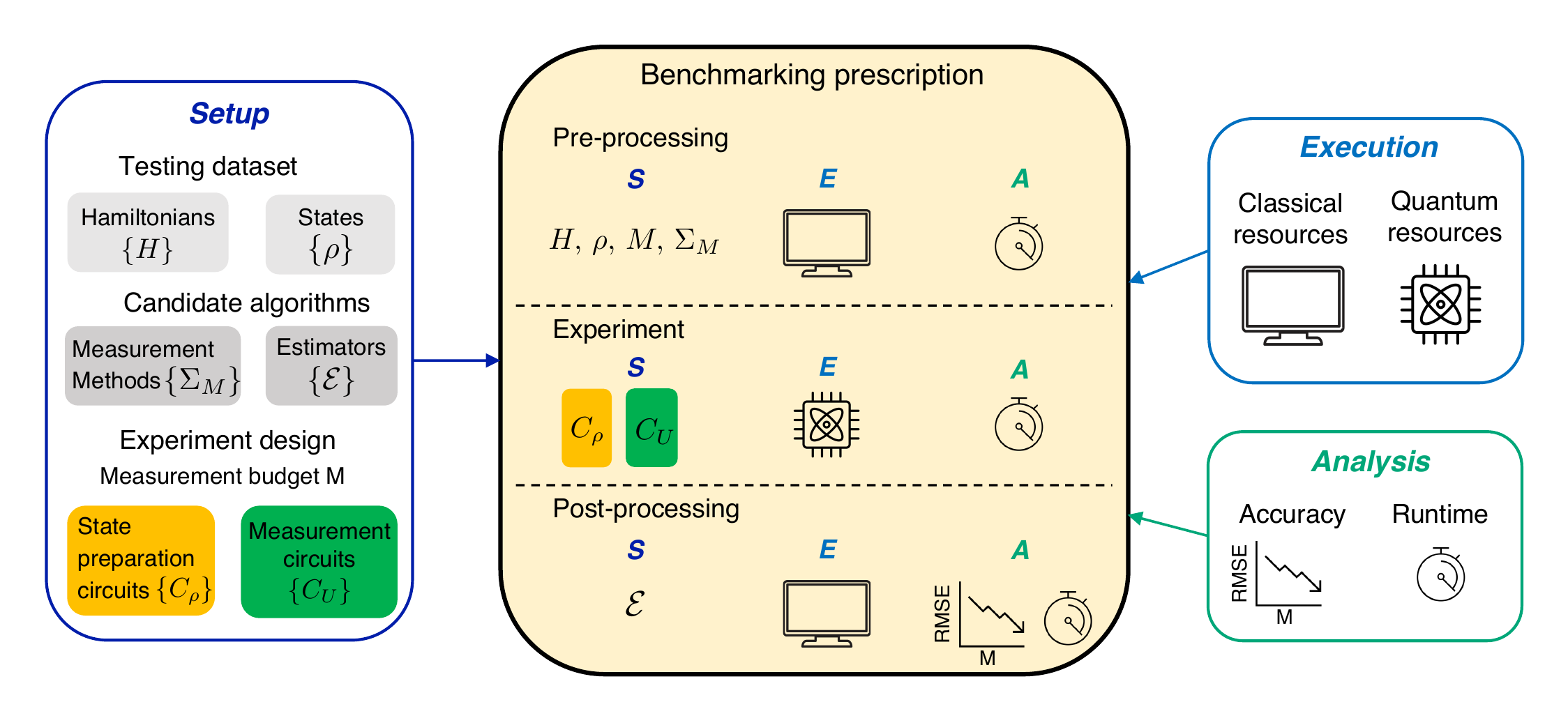}
\caption{Prescription for benchmarking a candidate measurement protocol (measurement method $\Sigma_M$ and estimator $\mathcal{E}$) on a Hamiltonian $H$ and state $\rho$ against measurement budget $M$. The benchmarking stages of setup (S), execution (E) and analysis (A) are shown for each step of a general measurement protocol. General measurement protocols, to be discussed in Section~\ref{subsec:general_protocol_meas}, can be divided into steps of classical pre-processing where measurement samples are generated, experiments on a quantum device and classical post-processing on data acquired. Each step has a benchmarking stage associated with it. As shown (left), setup defines the testing dataset, candidate algorithms and experiment design. Execution (right) defines the computational resources available for any step. Analysis (right) defines the metrics associated with each step.}
\label{fig:pep_sea_viz}
\end{figure}

As part of benchmarking different measurement protocols on a given $H$ and $\rho$, it will be convenient to break up each measurement protocol into the steps of pre-processing where the Pauli bases are generated using a measurement method, experiments where a state is measured in these bases on a quantum device and post-processing where an estimate is obtained from the data acquired using an estimator. In the rest of this paper, we will complete the benchmarking process for each of these steps in \benchname~ as commonly done in machine learning and computational chemistry benchmarks, consisting of the following stages: (i) \textit{setup}, (ii) \textit{execution}, and (ii) \textit{analysis}. An illustration of this benchmarking process (or prescription) is depicted in Figure~\ref{fig:pep_sea_viz}. (i) \textit{Setup} involves defining the test dataset of Hamiltonians and states under consideration, candidate measurement methods, estimators and the benchmarking experiment design. (ii) \textit{Execution} defines the computational resources available for executing the step such as classical computing (e.g., CPU, distributed computing, parallelization, simulator, etc.) and quantum computing (e.g., QPU, modular quantum devices etc.). (iii) \textit{Analysis} defines the performance metrics and the methods (e.g., empirical, inferential, etc.) to evaluate the performance metrics. The main distinction of \benchname~ from classical computational chemistry benchmarks is the availability of quantum devices for execution and as highlighted in the paragraph before, utilization of this resource will be important to account for. Finally, it is desirable that a benchmark is reproducible and reflects reality (or the performance obtained in experiment on quantum hardware). We demonstrate this by including data and analysis of \benchname~from experiments on IBM quantum devices. % Reproducibility, Invariance
%%%%%%%%%%%%%%

% Outline of paper
This paper is organized as follows. In Section~\ref{sec:background}, we first formalize the problem of estimating $\Tr(\rho H)$ for a given $n$-qubit Hamiltonian and access to an unknown $n$-qubit quantum state $\rho$ under the constraints of measuring in the Pauli basis. We then describe the benchmarking strategy followed in \benchname~in Section~\ref{subsec:CHSCQCBench_strategy} and then describe its setup, execution and analysis in the context of a general measurement protocol for estimating $\Tr(\rho H)$ in Section~\ref{subsec:general_protocol_meas}. This allows to kick off our discussion of different estimators that may be employed with various measurement methods in Section~\ref{sec:estimators}. In Section~\ref{sec:meas_methods}, we describe randomized and derandomized measurement methods using the framework of decision diagrams. In Section~\ref{sec:CHSCQCBench}, we formally describe the data set of molecular Hamiltonians and states considered as part of \benchname~before presenting the experimental protocol we follow. In Section~\ref{sec:results}, we report our results from \benchname~on the convergence behavior and resource utilization of various measurement methods. Finally in Section~\ref{sec:conclusion}, we comment on our benchmarking results and possible extensions of this work.

%%%%%%%%%%%%%%

\section{Background}\label{sec:background}
In this section, we introduce the problem of observable estimation, i.e., estimating $\Tr(\rho H)$ via Pauli measurements, given an $n$-qubit quantum Hamiltonian (or observable) $H$ and an $n$-qubit quantum state $\rho$. This is followed by the description of our benchmarking strategy for the observable estimation problem in Section~\ref{subsec:CHSCQCBench_strategy}. We then describe the different steps of setup, execution and analysis of \benchname~through a presentation of the general measurement protocol for estimating $\Tr(\rho H)$. 

A formal description of \benchname~is presented in Section~\ref{sec:CHSCQCBench} after describing the measurement methods in Section~\ref{sec:meas_methods} and estimators in Section~\ref{sec:estimators}. We now begin by introducing relevant notation.

\subsection{Notation}
We will denote the set of $n$-qubit Pauli operators as $\mathcal{P}_n = \{I,X,Y,Z\}^{\otimes n}$, the set of $n$-fold tensor products of the single-qubit Pauli matrices $\{I,X,Y,Z\}$. At times, it will be convenient to consider the set of $n$-fold tensor products of non-identity single-qubit Pauli matrices, which we denote by $\Omega_n = \{X,Y,Z\}^n$. For any $n$-qubit Pauli operator $Q$, we refer to its action on the $j$th qubit as $Q_j$ and hence have $Q = \bigotimes_{j=1}^n Q_j$. We denote the support of a Pauli operator as $\mathrm{supp}(Q) = \{j|Q_j \neq I\}$ and its weight as $\mathrm{wt}(Q) = |\mathrm{supp}(Q)|$.

We say that the $n$-qubit Pauli operator $B$ covers $n$-qubit Pauli operator $Q$ (or $Q$ is covered by $B$) if $Q$ can be obtained from $B$ by replacing some of the local Pauli matrices on single-qubits with identity. We then write $Q \triangleright B$.
We extend the same notation to sets of Pauli operators on the left hand side, e.g., $S\triangleright B$ if and only if all Pauli operators in $S$ are covered by $B$.
For example, $\{XXI,IXX,XIX\} \triangleright XXX$ but $\{ZZI,IZZ,ZIZ\} \not \triangleright XXX$.

\subsection{Observable estimation: Learning problem of measuring quantum Hamiltonians} \label{sec:learning_problem}
Consider an $n$-qubit Hamiltonian decomposed as a linear combination of $L$ Pauli terms
\begin{equation}
    H = \sum_{j=1}^{L} \alpha_j Q^{(j)}
    \label{eq:observable_defn}
\end{equation}
where $Q^{(j)}\in\mathcal{P}_n$ are $n$-qubit Pauli operators and  $\alpha_j \in \mathbb{R}$ are the corresponding coefficients.
We call the set $\mathbf{Q}\coloneqq\{Q^{(j)}\}_{j \in [L]}$ the \textit{target} observables where we used the notation $[L] = \{1,2,...,L\}$.

The \textit{observable expectation problem} is then as follows. Given an $n$-qubit quantum state $\rho$ (prepared by some quantum circuit), the goal is to estimate $E\coloneqq\Tr(\rho H)$ within error $\epsilon \in (0,1/2)$ using as few prepare-and-measure repetitions as possible. Note that $H$ can represent any physical observable; an typical example would be the qubit representation of a molecular Hamiltonian, as studied in the earliest papers to consider grouping of commuting Pauli measurements~\cite{kandala2017hardware,hempel2018trappedion}. In the process of obtaining an estimate $\hat{E}$ of $E$, we will obtain estimates of the $\Tr(\rho Q^{(j)})$, which will be denoted by $\hat{\omega}^{(j)}$. The true value of $\Tr(\rho Q^{(j)})$ will be denoted by $\omega^{(j)}$.

The main constraint that we will impose on our learning problem is that once $\rho$ has been prepared on a quantum device, we are only allowed to use measurements corresponding to $n$-qubit Pauli operators to learn values of $\Tr(\rho Q^{(j)})$ and hence $\Tr(\rho H)$. This ensures that we do not have any access to quantum resources such as entanglement for learning, and our measurement circuits are composed of single-qubit operators. As discussed above, this is a reasonable constraint to impose on existing and near-term noisy quantum hardware where one would want to prioritize depth in the state preparation circuit over depth in the measurement circuit.

\subsection{Strategy for \benchname}\label{subsec:CHSCQCBench_strategy}
The goal of \benchname~is to assess the performance of and rank different measurement methods in estimating $\Tr(\rho H)$ through local Pauli measurements on a quantum computer, for $n$-qubit Hamiltonians $H$ and $n$-qubit quantum states $\rho$ prepared on a quantum computer.
% The ``common quantum computation'' is then that of observable estimation or estimating $\Tr(\rho H)$.
To start the description of the benchmarking setup, we need to define the set of objects, i.e., types of Hamiltonians $H$ and quantum states $\rho$ considered as part of the test suite for our candidate measurement methods. 

Analogous to classical benchmarks of MS COCO \cite{lin2014coco} and on QM datasets~\cite{blum2009qmdata,ruddigkeit2012qm9data} (described in Section~\ref{sec:intro}), we consider the broader learning context of these measurement methods when used in near-term hybrid quantum-classical algorithms, which is typically ground state estimation. We thus consider objects from this natural context. The set of Hamiltonians considered here include small molecular Hamiltonians of varying sizes and with varying Pauli weight distributions. We consider different types of states that would be expected during the runtime of a hybrid quantum-classical algorithm such as VQE. For example, we benchmark against the Hartree-Fock (HF) state, which is classically simulatable and a possible initial state for many ground estimation algorithms. During the course of a (successful) VQE run, one could also expect to see an approximate ground state at the very end, but for the purpose of benchmarking such states are not desirable since they may be difficult to prepare. Instead, we benchmark against quasi-random states prepared by a typical low-depth ansatz with random parameter settings. These random states serve as a proxy for typical intermediate states obtained in the middle of a VQE optimization, since although in that case the parameter setting would not be random, it would not in general bear any particular relation to the Pauli decomposition of the target observable. The overall code base of \benchname~is designed such that any new Hamiltonian can be easily added to the existing dataset and measurement methods benchmarked against it.

% Performance metrics
The most popular metric used so far is that of accuracy, i.e., root mean square error (RMSE) in the estimate of $\Tr(\rho H)$ for a given budget of measurements. However, a highly accurate measurement method may not be useful in practice as the classical computational runtime required for set up or optimization may be prohibitive and the quantum resources required too demanding. It is thus imperative to analyze the resources utilized in obtaining an accurate estimate through a measurement method. We further stress that we need to account for both classical and quantum resources as the subroutine of obtaining expectation values with respect to different quantum observables is inherently hybrid quantum-classical in nature, requiring different experiments to be executed on the quantum device and classical post-processing of the measurements in addition to pre-processing to decide the experiments themselves. 

In the next section, we discuss a general measurement protocol and comment on resource utilization in the different steps of the protocol.

\subsection{General measurement protocol}\label{subsec:general_protocol_meas}
\begin{figure}[ht!]
\centering
\includegraphics[width=0.9\textwidth]{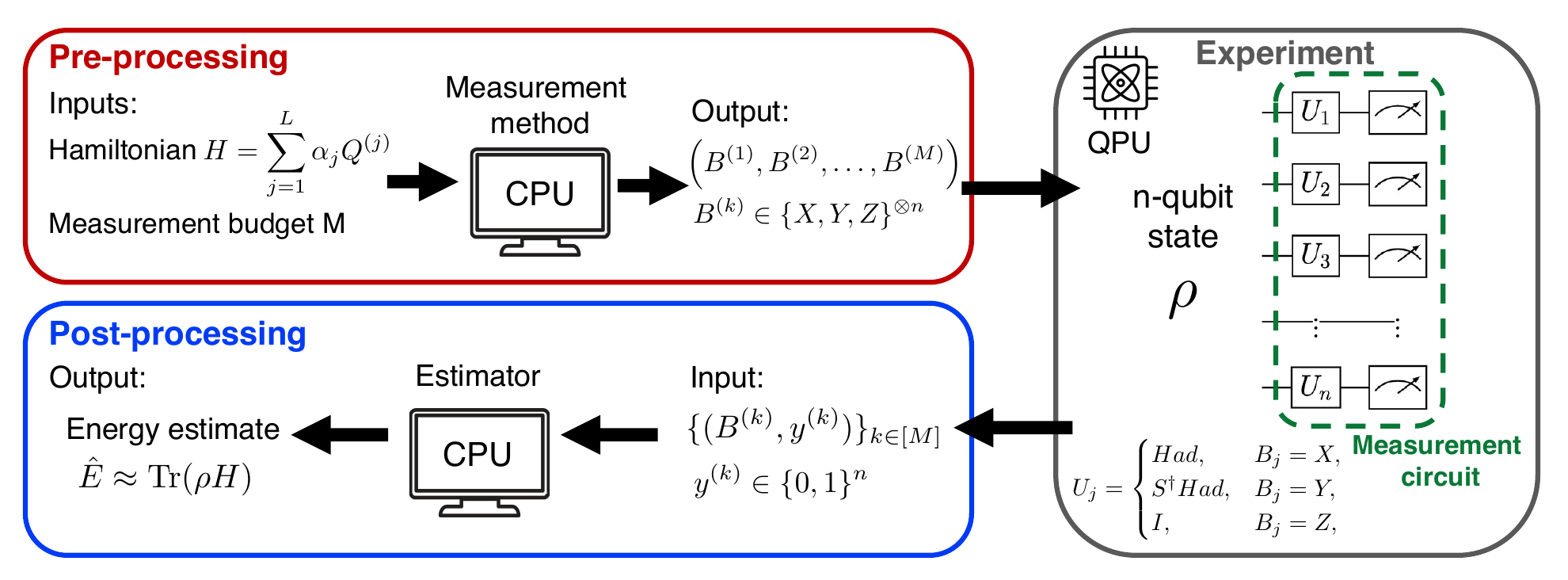}
\caption{Schematic of estimation of $\Tr(\rho H)$: The procedure is divided into three steps of (i) pre-processing on a classical computer (CPU), (ii) experiments on a quantum device or quantum processing unit (QPU), and (iii) post-processing on the CPU. In (i), the measurement method plays the central role of deciding which measurement bases (denoted by $B$ here) to execute on the QPU given inputs of an $n$-qubit Hamiltonian and measurement budget $M$. In (ii), experiments are executed on the QPU using the inputs of the Pauli measurement bases output by step (i). A measurement circuit corresponding to an arbitrary basis $B$ is shown inset ($Had$ denotes the Hadamard gate, and $S^\dagger$ denotes the inverse phase gate, which are used to transform the local measurement basis). Finally, in (iii) estimation is carried out on measurement results of the form $(B,y)$ where $B$ are the Pauli measurement bases and $y$ are the corresponding measurement outcomes from the QPU.}
\label{fig:flow_randomized_measurements}
\end{figure}
In this section, we describe the general procedure along with resource utilization for the problem of estimating $\Tr(\rho H)$ on $n$ qubits, given measurement budget $M$. The measurement budget is equivalent to the number of total shots we are allowed gather from the quantum device or the number of times the device is queried. 

% Broadly, we consider the steps of pre-processing, experiment and post-processing
The general procedure is schematically depicted in Figure~\ref{fig:flow_randomized_measurements} and involves three steps: (i) pre-processing on a classical computer (CPU), (ii) experiments on a quantum device (or QPU for quantum processing unit), and (iii) further post-processing of data acquired from the quantum device on a classical computer. We now describe each of these steps in detail. We will also explicitly state the benchmarking setup, execution, and analysis associated with each step. First, we describe the experiments executed on the quantum device as this decides the formulation of the pre-processing and post-processing steps.

\paragraph*{Experiment.} Each experiment on the quantum device involves the preparation of the $n$-qubit quantum state of interest followed by a measurement circuit. In an arbitrary step of VQE, this state would correspond to a parametrized quantum circuit or an ansatz with a certain set of assigned parameters. In quantum Krylov methods, this state may correspond to a certain time-evolved state. After the state is prepared, a measurement circuit is applied which involves application of single-qubit unitaries $\otimes_{i=1}^n U_i$ followed by a measurement in the computational basis. We denote the outcome of a measurement which is an $n$-bit binary string as $y \in \{0,1\}^n$. For any arbitrary qubit $j$, the single-qubit unitary $U_j$ corresponds to measuring qubit $j$ in a non-trivial Pauli basis
\begin{equation}
    U_j = \begin{cases} Had, & B_j=X, \\ S^\dagger Had, & B_j=Y, \\ I, & B_j=Z, \end{cases}    
\end{equation}
where we have denoted the $n$-qubit Pauli basis as $B$, the subscript $j$ denotes the Pauli matrix on qubit $j$, $Had$ is the Hadamard gate, and $S=\mathrm{diag}(1, i)$ is the phase gate. As discussed earlier in Section~\ref{sec:intro}, we restrict measurement circuits to involve measurements in the Pauli basis due to depth limitations on currently available noisy quantum hardware and hence the preference for shallow measurement circuits. In summary, the input to the experiment step for a measurement budget $M$ is a set of $M$ Pauli measurement bases $\{B^{(k)}\}_{k \in [M]}$ corresponding to the measurement circuits executed on the quantum device and the output from this step is a set of measurement outcomes with each one corresponding to a measurement basis $\{(B^{(k)}, y^{(k)})\}_{k \in [M]}$. This output is then later used in the post-processing step for obtaining an estimate of $\Tr(\rho H)$. This completes the description of the experiment step. 

The benchmarking setup includes defining the state preparation circuit (either after compiling a classical description of a digital quantum circuit to be implemented or setting the parameters in a parameterized quantum circuit), and the measurement circuits to be used for the chosen measurement bases. As part of the benchmarking execution, we need to define the type of quantum computing resource being utilized (e.g., QPU, modular, etc.) as well as any classical computing resource (e.g., CPU, distributed, parallel, etc.) utilized for compilation and error mitigation. Finally, as part of the benchmarking analysis, this step involves the quantum wall-clock time over the experiments executed and the metric of computational runtime associated with compilation or error mitigation.

% Pre-processing step: Introduce some notation and distinguish between have access to query distributions versus just being able to sample from an unknown distribution
% TODO: More discussion in pre-processing step required
\paragraph*{Pre-processing.} In this step, a measurement method is used to propose a set of Pauli measurement bases $\{B^{(k)}\}_{k \in [M]}$ that is inputted to the experiment step. The inputs to the measurement method are the $n$-qubit Hamiltonian $H$, measurement budget $M$, and any available prior information of the quantum state $\rho$. Taking the measurement basis operator on any particular qubit to be the identity $I$ corresponds to not measuring that qubit, and hence does not reveal any information about a target observable $Q$. Therefore, we consider the alphabet of measurement bases as $\mathcal{Q} = \{X,Y,Z\}^{\otimes n}$ and call it the query space. We denote a distribution over $\mathcal{Q}$ as $\beta$ and call it the query distribution. The probability mass associated with a Pauli operator $P \in \mathcal{Q}$ is given by $\beta(P)$.

The benchmarking setup includes defining the Hamiltonian $H$ under consideration, quantum state $\rho$ to be measured (either through a classical description of the state preparation circuit required to be implemented or parameters associated with a parameterized quantum circuit), and measurement method $\Sigma_S$ to be used for generating the samples. As part of the benchmarking execution, we need to define the type of computational resource being utilized (e.g., CPU, distributed computing, parallelization, etc.). Finally, as part of the benchmarking analysis, this step only involves the metric of computational runtime associated with the measurement method generating $M$ samples.

% Post-processing step: Give some more notation for eigenvalues measurements
% Describe empirical estimate of Tr(B rho)
% How measurements from B can also be used for other Paulis Q which are covered by B
\paragraph*{Post-processing.} After measurement outcomes~$\{(B^{(k)}, y^{(k)})\}_{k \in [M]}$ against $M$ Pauli measurement bases are acquired in the experiment step, they are passed on to an estimator $\mathcal{E}$ in the post-processing step. Suppose there is a Pauli measurement basis $B$ which is queried $m_B$ times and the corresponding measurement outcomes are $\{y_B^{(k)}\}_{k \in [m_B]}$. We can then compute an estimate, which we denote by $\hat{\omega}(B)$, of $\Tr(\rho B)$ as follows:
\begin{equation}
    \hat{\omega}(B) = \frac{1}{m_B} \sum \limits_{k=1}^{m_B} \mu^{(k)}(B)  = \frac{1}{m_B} \sum \limits_{k=1}^{m_B} \prod_{j=1}^n (-1)^{y_j^{(k)}},
    \label{eq:empirical_estimate_trace_BH}
\end{equation}
where $\mu^{(k)}(B)=\prod_{j=1}^n (-1)^{y_j^{(k)}}$ is eigenvalue measurement of $\rho$ in the basis $B$ corresponding to the outcome $y^{(k)}$. But, it turns out we can do even more with these measurements. Suppose there is a target Pauli term $Q$ in the decomposition of the Hamiltonian $H$ (Eq.~\ref{eq:observable_defn}) which is covered by $B$, i.e., all the non-trivial single-qubit Pauli matrices in the tensor product of $Q$ coincide with those in $B$. The eigenvalue measurement of $\rho$ in basis $Q$ could then be obtained from the measurement outcomes of measuring $\rho$ in basis $B$.

Let us make this more concrete by introducing some relevant notation. For a full weight Pauli operator $B$, we let $\mu(B,j) = (-1)^{y_j}$ denote the eigenvalue measurement when qubit $j$ is measured in the basis $B_j$ and as corresponding to measurement outcome $y$. For for a subset $A \subset [n]$, we define $\mu(B,A) := \prod_{j \in A} \mu(B,i) = \prod_{j \in A} (-1)^{y_j}$ with the convention that $\mu(B,\emptyset)=1$. The eigenvalue measurement of $\rho$ in basis $Q$ corresponding to measurement outcome $y^{(k)}$ of measuring $\rho$ in basis $B$ is then given by $\mu^{(k)}(B,\mathrm{supp}(Q))$. We thus note that eigenvalue measurements of $\rho$ in basis $B$ can then be used to obtain estimates of not only $\Tr(\rho B)$ but also of $\Tr(\rho Q)$ for all $Q$ that are covered by $B$. These estimates can then be combined to give an estimate $\hat{E}(\rho)$ of $\Tr(\rho H)$. The goal of the estimator $\mathcal{E}$ is to do this in a computationally efficient fashion while using the available measurements to come up with an accurate estimate. Here, we have hinted at the Monte-Carlo estimator \cite{robert1999monte}. In Section~\ref{sec:estimators}, we will give an overview of different estimators that can be used in the post-processing step.

The benchmarking setup includes defining the estimator $\mathcal{E}$ used. As part of the benchmarking execution, we need to define the type of computational resource being utilized (e.g., CPU, distributed computing, parallelization, etc.). Finally, the benchmarking analysis involves the accuracy metric of the output learning error (RMSE) and performance metric of computational runtime associated with running the estimator on the acquired data.

% Algorithm
A sequential algorithmic viewpoint of the general protocol for observable estimation discussed thus so far is presented in Algorithm~\ref{algo:energy_estimation}. While benchmarking different measurement methods, it will also be convenient to define the PEP-SEA matrix, used as an abbreviation of (P)re-processing (E)xperiment (P)ost-processing - (S)etup (E)xecution (A)nalysis matrix, that summarizes the different stages of benchmarking for each step of the general measurement protocol. The PEP-SEA matrix is summarized in Table~\ref{tab:PEP-SEA_matrix}.

We have noted that the design of estimators and measurement methods can be interdependent. For example, a trivial estimator could be designed to estimate $\Tr(\rho B)$ for a measurement basis $B$ that also occurs as a target Pauli term in the Hamiltonian $H$ using the corresponding measurement outcomes but not use the same measurement outcomes to estimate $\Tr(\rho Q)$ for a Pauli $Q$ that is covered by $B$. This would put a severe restriction on sets of useful measurement bases to the experiment and limit the flexibility of the measurement methods. In this work, we only discuss estimators which are designed with compatibility of different Pauli operators in mind, and discuss which of those estimators may be equipped with the various measurement methods presented here.

In Section~\ref{sec:estimators}, we describe different estimators that can be used in post-processing. In Section~\ref{sec:meas_methods}, we discuss various measurement methods that either construct a query distribution $\beta$ in order to sample measurement bases $B$ from it or use a routine to sample measurement bases $B$ without direct access to the underlying distribution. We will also comment on the compatibility of different measurement methods with different estimators in Section~\ref{sec:meas_methods}. 
\begin{algorithm}[H]
    \caption{Estimation of $\Tr(\rho H)$ through different measurement methods} \label{algo:energy_estimation}
    \textbf{Input}: Measurement budget $M$, Hamiltonian $H$, Measurement Method $\mathrm{\Sigma}_S$, Estimator $\mathcal{E}$
    \begin{algorithmic}[1]
        \For{sample $m=1:M$}
            \State Generate measurement basis $B \in \{X,Y,Z\}^{\otimes n}$ using $\mathrm{\Sigma}_S$
            \For{qubit $k=1:n$}
                \State Measure qubit $k$ in basis $B_k$
                \State Save eigenvalue measurement $\mu(B,k) \in \{-1,+1\}$
            \EndFor
            \State Update observable expectation $\hat{E}$ using estimator $\mathcal{E}$ on acquired eigenvalue measurements 
        \EndFor 
    \end{algorithmic}
    \textbf{Output}: $\hat{E}$
\end{algorithm}

\begin{table}[H]
\small
\centering
{\renewcommand{\arraystretch}{1.2}
\begin{tabular}{r|l|l|l|}
\cline{2-4}
\multicolumn{1}{l|}{} & \multicolumn{1}{c|}{\textbf{Setup (S)}} & 
\multicolumn{1}{c|}{\textbf{Execution (E)}} & 
\multicolumn{1}{c|}{\textbf{Analysis (A)}}  \\ \hline
\multicolumn{1}{|r|}{\textbf{Pre-processing (P)}} & 
\begin{tabular}[c]{@{}l@{}}
    Meas. budget: $M$ \\
    Objects: $H$, $\rho$\\ 
    Meas. Method: $\Sigma_S$
\end{tabular}  & 
\begin{tabular}[c]{@{}l@{}}
    Classical computational\\ 
    resource utilized\\ (CPU, distributed, parallel)
\end{tabular} & 
    Classical runtime \\ \hline
\multicolumn{1}{|r|}{\textbf{Experiment (E)}} 
& \begin{tabular}[c]{@{}l@{}}
    State preparation circuit for $\rho$\\ 
    Meas. circuit: $\{U_j\}_{j \in [n]}$ \\
    from $\{B_j\}_{j \in [n]}$
\end{tabular} & 
\begin{tabular}[c]{@{}l@{}}
    Quantum computing\\ 
    resource utilized\\ 
    (QPU, modular)
\end{tabular}  & 
\begin{tabular}[c]{@{}l@{}}
    Quantum wall-clock time\\ 
    Classical wall-clock time\\ 
    (error mitigation, compilation)\\ 
    Quantum coherence time
\end{tabular} \\ \hline
\multicolumn{1}{|r|}{\textbf{Post-processing (P)}} & 
Estimator: $\mathcal{E}$ & 
\begin{tabular}[c]{@{}l@{}}
    Classical computational\\ 
    resource utilized\\ 
    (CPU, distributed, parallel)
\end{tabular} & 
\begin{tabular}[c]{@{}l@{}}
    Classical runtime\\ 
    Output accuracy (RMSE)
\end{tabular} \\ \hline
\end{tabular}
}
\caption{Definition of the PEP-SEA matrix.}
\label{tab:PEP-SEA_matrix}
\end{table}

\subsection{Summary of performance metrics in \benchname}\label{subsec:CHSCQCBench_metrics_design}
To account for computational runtime in addition to the convergence of the algorithm at hand, one common approach used across different classical computational chemistry packages is to measure the wall-clock time required to achieve some threshold accuracy. As the learning task of estimating $\Tr(\rho H)$ is a subroutine in many hybrid quantum-classical algorithms, we need to account for the computational time spent on the quantum device, computational time spent on the classical device, and any latencies in between the quantum and classical hardware, to measure the overall wall-clock time. We refer to the total time spent on a quantum device as quantum wall-clock time and this includes time duration associated with experiment execution, measurements and resets. We refer to the total time spent on a classical device as classical wall-clock time, which includes setting up different measurement methods (e.g., optimization), post-processing of measurements (e.g., estimation), and compilation of quantum circuits to the native set of gates of the quantum hardware. % \tcred{Time cost associated with classical electronics such as microwave generator and other latencies?}

A performance metric for resource utilization could thus simply be the sum of the quantum wall-clock time and classical wall-clock time for a measurement method in reaching a cutoff of accuracy. However, quantum computers are not yet a mature technology compared to classical computers. This suggests that a stronger approach to benchmarking would be to allow some flexibility in weighting the quantum and classical costs, since wall-clock time may not all be equivalent across the classical and quantum phases of the experiment.

Towards this end, we propose a heuristic in Section~\ref{sec:results} to rank different measurement methods based on a weighted-sum of the quantum wall-clock time and classical wall-clock time to reach an specified cutoff of chemical accuracy in RMSE of the resulting energy estimates. We expect quantum devices to progress rapidly over the next few year, and so for the benchmarks to be robust to this progress, the weights should vary with time and be revised with new advances. A functional form of how these weights may change with time may be useful but is outside the scope of the current paper.

\section{Estimators}\label{sec:estimators}
In this section we discuss different estimators $\mathcal{E}$, which take center stage in the post-processing step of the general protocol for estimating $\Tr(\rho H)$ (Figure~\ref{fig:flow_randomized_measurements}) and that can be used alongside measurement methods as shown in Algorithm~\ref{algo:energy_estimation}.
Suppose we generate a set of $M$ Pauli measurement bases through a specified measurement method.
We will denote this set of measurements by $\mathbf{B} = \{B^{(s)}\}_{s \in [M]}$ where $B^{(s)} \in \Omega_n = \{X,Y,Z\}^{\otimes n}\, \forall s \in [M]$.
Recall that we denote the corresponding measurement outcomes as $\mu(B^{(s)}) \in \{-1,+1\}^{\otimes n}$ with $\mu(B^{(s)},j)$ the measurement outcome on qubit $j$.

The goal is to use the $M$ examples of $\{(B^{(s)},\mu(B^{(s)}))\}_{s \in [M]}$ to estimate $\Tr(\rho H)$. We do this by obtaining estimates of $\Tr(\rho Q^{(j)})$ which we denote by $\hat{\omega}^{(j)}$, using one of three possible estimators: a Monte Carlo (MC) estimator, a weighted Monte Carlo (WMC) estimator, and a Bayesian estimator, which we introduce in the next subsection. Given the $\hat{\omega}^{(j)}$, an estimate of $\Tr(\rho H)$ is obtained as
\begin{equation}
    \hat{E}_G = \sum \limits_{j=1}^L \alpha_j \hat{\omega}^{(j)}.
    \label{eq:energy_estimate}
\end{equation}

\subsection{Monte-Carlo Estimator}\label{subsec:mc_estimator}
Let us first define the \textit{hit} function 
\begin{equation}
    h(Q^{(j)};\mathbf{B}) = \sum \limits_{s=1}^M \mathds{1}\{ Q^{(j)} \triangleright B^{(s)}\}
    \label{eq:hit_function_basis}
\end{equation}
which counts the number of Pauli measurement bases that cover (hit) $Q^{(j)}$. 
The MC estimate of $\hat{\omega}^{(j)}$ is then simply given by 
\begin{equation}
    \hat{\omega}^{(j)} = 
    \begin{cases}
    & \frac{1}{h(Q^{(j)};\mathbf{B})} \sum_{s=1}^M \mathds{1}\{ Q^{(j)} \triangleright B^{(s)}\} \mu \left(B^{(s)}, \mathrm{supp}(Q^{(j)})\right), \quad h(Q^{(j)};\mathbf{B}) \geq 1\\
    & 0, \quad h(Q^{(j)};\mathbf{B}) = 0.
    \end{cases}
    \label{eq:mc_estimator}
\end{equation}
This is merely the empirical average of the measurements corresponding to $Q^{(j)}$. As the MC estimator does not require access to a query distribution $\beta$, it can be used with any measurement methods, including those that generate measurement bases without giving us direct access to an underlying query distribution. Finally, it can be readily verified that the MC estimator is an unbiased estimator.

% Need to work on the motivation for Laplace correction further
\paragraph{Laplace smoothing.}\label{sec:laplace_smoothing} In any collected dataset $D = \{(B^{(s)},\mu(B^{(s)}))\}_{s \in [M]}$, there is a possibility that there are no measurements covering an observable $Q^{(\ell)}$ for some $\ell \in [L]$ and thus the hit $h(Q^{(\ell)};\mathbf{B})=0$. This happens when the query distribution $\beta$ assigns very low probability to measurement bases covering $Q^{(\ell)}$. For example, this occurs when the measurement method is based on $L_1$ sampling and the corresponding coefficient $\alpha_\ell$ in the decomposition of $H$ has a low magnitude relative to the $1$-norm of the coefficients $\sum_{j=1}^L |\alpha_j|$.

Now suppose that in this situation where the probability of a measurement basis covering $Q^{(\ell)}$ is very low, we do get lucky and obtain a single measurement of it.
% Prior to this measurement being obtained, the empirical probability of the single shot measurement corresponding to $Q^{(\ell)}$, denoted by $\lambda^{(\ell)}$, being $+1$ or $-1$ is then set to zero.
When this occurs the estimate $\hat{\omega}^{(\ell)}$ will jump from the value $0$ (prior to that shot) to the measurement outcome, either $+1$ or $-1$.
% This results in $\hat{\omega}^{(\ell)}$ being set to a value of $1$ or $-1$.
This may result in large contribution to the uncertainty in the overall estimate of $\Tr(\rho H)$ from $\alpha_\ell \hat{\omega}^{(\ell)}$ even when $\alpha_\ell$ is small, if the true value of $Q^{(\ell)}$ far from the obtained measurement outcome (as for example is guaranteed if the true value is close to 0).
To avoid this, we artificially adjust the empirical probabilities via Laplace smoothing (also called additive smoothing)~\cite{manning2008introduction} as follows
\begin{equation}
    P(\lambda^{(j)}=1) = \frac{m^{(j)}_0 + \gamma}{m^{(j)}_0 + m^{(j)}_1 + 2\gamma}, \quad P(\lambda^{(j)}=-1) = \frac{m^{(j)}_1 + \gamma}{m^{(j)}_0 + m^{(j)}_1 + 2\gamma},
\end{equation}
where $\gamma$ is the smoothing parameter and $m^{(j)}_0$ (or $m^{(j)}_1$) are the number of measurements in $D$ which cover $Q^{(j)}$ and correspond to an eigenvalue measurement of $+1$ (or $-1$). Formally, we have
\begin{equation}
    m^{(j)}_k = \frac{1}{2} \sum_{s=1}^M \mathds{1}\{ Q^{(j)} \triangleright B^{(s)}\} \left(1 + (-1)^k \mu \left(B^{(s)}, \mathrm{supp}(Q^{(j)})\right) \right), \, k \in \{0,1\}.
\end{equation}

The resulting MC estimate of $\hat{\omega}^{(j)}$ with Laplace smoothing is then
\begin{equation}
    \hat{\omega}^{(j)} = P(\lambda^{(j)}=1) - P(\lambda^{(j)}=-1) = \frac{1}{h_\gamma(Q^{(j)};\mathbf{B})} \sum_{s=1}^M \mathds{1}\{ Q^{(j)} \triangleright B^{(s)}\} \mu \left(B^{(s)}, \mathrm{supp}(Q^{(j)})\right),
    \label{eq:mc_estimator_laplace_smoothing}
\end{equation}
where $h_\gamma(Q^{(j)};\mathbf{B})$ is the smoothed hit function related to the original hit function as $h_\gamma(Q^{(j)};\mathbf{B}) = h(Q^{(j)};\mathbf{B}) + 2\gamma$. Note that for $\gamma=0$, we have no smoothing and re-obtain the original MC estimator (Eq.~\ref{eq:mc_estimator}). The value of $\gamma=1$ corresponds to the case when we assume the prior probability of $P(\lambda^{(j)}=1) = P(\lambda^{(j)}=-1) = 1/2$ are uniform. The value of $\gamma=0.5$ corresponds to the case when the prior probabilities are the Jeffrey's prior. Typically, $\gamma$ is set to a value in $(0,1)$ or is treated as a hyperparameter to be fine-tuned later.

\subsection{Weighted Monte-Carlo Estimator}\label{subsec:wmc_estimator}
In the MC estimator, all the eigenvalue measurements are weighted uniformly in computing $\hat{\omega}^{(j)}$. This can be modified by weighting the different samples non-uniformly as follows:
\begin{equation}
    \hat{\omega}^{(j)} = \frac{1}{M}\sum_{s=1}^M w^{(s)} \mu \left(B^{(s)}, \mathrm{supp}(Q^{(j)})\right),
\end{equation}
where we have introduced weights $\{w^{(s)}\}_{s \in [M]}$ that satisfy $\sum_s w^{(s)}=1$. This is often desirable to ensure stability of estimation~\cite{casella1996statistical}, for incorporating prior information or for the purpose of importance sampling when a proposal distribution is used instead of the query distribution $\beta$ for generating samples of measurement bases $B$.

Here we consider weights based on the query distribution, which can be interpreted as a self-normalization. Let the probability of a Pauli operator $Q$ being covered by a query distribution $\beta$ be denoted by $\xi(Q,\beta)$. This is equal to the probability with respect to $\beta$ of generating a sample measurement basis $B$ that covers $Q$:
\begin{align}
    \xi(Q,\beta) = \sum_{B \in \Omega_n} \mathds{1}\{ Q \triangleright B\} \beta(B),
    \label{eq:prob_of_coverage_Q}
\end{align}
where we have assumed that the alphabet of $\beta$ is $\Omega_n$. The resulting WMC estimator from setting the weights as $w^{(s)} = \mathds{1}\{ Q^{(j)} \triangleright B^{(s)}\}/{\xi(Q^{(j)},\beta)}$ (with $w^{(s)}$ set to $0$ if $\mathds{1}\{ Q^{(j)} \triangleright B^{(s)}\}=0$) is then given by
\begin{equation}
    \hat{\omega}^{(j)} = \frac{1}{M} \sum_{s=1}^M \frac{\mathds{1}\{ Q^{(j)} \triangleright B^{(s)}\}}{\xi(Q^{(j)},\beta)} \mu \left(B^{(s)}, \mathrm{supp}(Q^{(j)}) \right).
    \label{eq:weighted_mc_estimator}
\end{equation}
We note that the expectation of the hitting function $\Exp_\beta[h(Q^{(j)};\mathbf{B}] = M \xi(Q^{(j)},\beta)$. We can thus interpret the weighting in the WMC estimator as assigning a value to $\mathds{1}\{ Q^{(j)} \triangleright B^{(s)}\}/h(Q^{(j)};\mathbf{B})$ according to $\beta$ and not through the samples actually obtained, in contrast to the MC estimator of Eq.~\ref{eq:mc_estimator}. 

The WMC estimator can be used when the measurement method has access to the query distribution $\beta$ and we can evaluate $\beta(P)$ for an $n$-qubit Pauli operator $P$. In literature, WMC estimators have been used where $\beta$ is a product distribution \cite{huang2020predicting,hadfield2020measurements} as well as for more general cases \cite{hillmich2021decision}.
It is not immediately clear whether the MC or WMC estimator would be preferred in practice.
To answer this question, we compare the behavior of these estimators in Section~\ref{subsec:comparison_estimators} in combination with various applicable measurement methods.

\textit{Remark.} The MC estimator (Eq.~\ref{eq:mc_estimator}) and WMC estimator (Eq.~\ref{eq:weighted_mc_estimator}) may display different behaviors for low number of samples. However, it can be shown they are asymptotically equivalent.

\subsection{Bayesian Estimator}\label{subsec:bayesian_estimator}
We now take a Bayesian approach in obtaining estimates $\hat{\omega}^{(j)}$ of $\Tr(\rho Q^{(j)})$.
The estimates $\omega^{(j)}$ are obtained through single-shot measurements of $\rho$ in a Pauli basis $B$ that covers $Q^{(j)}$ (i.e., $Q^{(j)} \triangleright B$), which we denote by the random variable $\lambda^{(j)}$ which takes values in $\{+1,-1\}$.
Let the underlying probability of observing $\lambda^{(j)}$ to be $+1$ ($-1$) be $\theta^{(j)}_0$ ($\theta^{(j)}_1$).
% It is clear that $\theta^{(j)}_1 = 1 - \theta^{(j)}_0$.
We then model the random vector $\theta^{(j)} = (\theta^{(j)}_0, \theta^{(j)}_1)$ as a Dirichlet random variable of order $2$ and with hyperparameters $a = (a_1,a_2)$, i.e., $P(\theta^{(j)}) = \mathrm{Dir}(2,a)$.
We set $a = (1,1)$ which corresponds to a uniformly distributed prior.

Given a data set of $M$ measurements $D=\{(B^{(s)},\mu(B^{(s)}))\}_{s \in [M]}$ as follows, we can then update the probability distribution over $\theta^{(j)}$ as
\begin{equation}
    P(\theta^{(j)}|D) = \frac{P(D|\theta^{(j)})P(\theta^{(j)})}{P(D)}
    \label{eq:bayes_law}
\end{equation}
where we have denoted $P(\theta^{(j)})$ as the prior probability of $\theta^{(j)}$, $P(D|\theta^{(j)})$ as likelihood or conditional probability of obtaining the measurements in $D$ given $\theta^{(j)}$, $P(D)$ as the evidence, and $P(\theta^{(j)}|D)$ as the posterior probability of $\theta^{(j)}$ given $D$.

Typically, computing the posterior distribution through Bayes law (Eq.~\ref{eq:bayes_law}) is expensive and one needs to resort to alternate approximate methods such as Markov Chain Monte Carlo sampling and particle filter methods~\cite{sarkka2013bayesian}. However, in this case, the prior $P(\theta^{(j)})$ is a conjugate prior to the likelihood distribution, which is given by
\begin{equation}
    P(D|\theta^{(j)}) = \left(\theta^{(j)}_0\right)^{m_0}\left(\theta^{(j)}_1\right)^{m_1}, \quad m_k = \frac{1}{2} \sum_{s \in [M]} \mathds{1}\{B^{(s)} \covers Q^{(j)}\} \left(1 + (-1)^k \mu \left(B^{(s)}, \mathrm{supp}(Q^{(j)})\right)\right), k \in \{0,1\}.
\end{equation}
This means that our prior $P(\theta^{(j)})$ and posterior $P(\theta^{(j)}|D)$ both belong to the same family of distributions, which in this case are Dirichlet distributions. The posterior is simply given by $P(\theta^{(j)}|D) = \mathrm{Dir}(2,a+m)$ where $m = (m_0,m_1)$. Ultimately, the posterior distribution is given by 
\begin{equation}
    P(\theta^{(j)}|D) = \frac{ \left(\theta^{(j)}_0\right)^{m_0}\left(\theta^{(j)}_1\right)^{m_1} }{B(m_0 + a_0, m_1 + a_1)}
    \label{eq:posterior_distrn}
\end{equation}
where $B(k_1,k_2) = \Gamma(k_1)\Gamma(k_2)/\Gamma(k_1 + k_2)$ and $\Gamma(z) = \int_{0}^\infty x^{z-1}\exp(-x)dx$ is the gamma function. In practice, we may receive datasets sequentially and the Bayes law (Eq.~\ref{eq:bayes_law}) is then computed sequentially with the posterior becoming the prior for the next dataset. Finally, an estimate of the expected value of $\omega^{(j)}$ or its variance can be obtained as
\begin{align}
    \mathbb{E}[\omega^{(j)}] &= 2 \mathbb{E}[\theta^{(j)}_0] - 1 = \frac{m_0 - m_1}{m_0 + m_1 +2} \\
    \Var[\omega^{(j)}] &= 4 \Exp[ \theta^{(j)}_0(1-\theta^{(j)}_0) ] = 4 \frac{ (m_0 + 1)(m_1 + 1)}{(m_0 + m_1 +2)(m_0 + m_1 +3)}
    \label{eq:analytical_expressions_estimates}
\end{align}

Thus, for carrying out Bayesian estimation of $\omega^{(j)}$, it is enough to keep track of the cumulative number of shots corresponding to measurements of $Q^{(j)}$ yielding values of $+1$ and $-1$. We also observe that the expected value of the estimate $\omega^{(j)}$ coincides with our MC estimate (Eq.~) with Laplace error correction of $\gamma = 1$. This is not pure coincidence as Laplace error correction can be motivated through Bayesian estimation.

So far, we have shown how to carry out Bayesian estimation of $\omega^{(j)}$, but in many instances other quantities such as the covariance of estimates of two different target Pauli operators may be of interest.
For a discussion on Bayesian estimation in this latter case, we refer the reader to \cite[Appendix~B]{shlosberg2023adaptiveestimation}.

\section{Measurement Methods}\label{sec:meas_methods}
In this section, we give an overview of different measurement methods $\Sigma_S$ that can be used in the pre-processing step of the general protocol (Section~\ref{subsec:general_protocol_meas}). We will primarily focus on randomized measurements where, as the name suggests, there is randomization involved in producing the measurement bases.
We also consider how these methods may be derandomized.
Among randomized measurements, there are some methods that involve the explicit construction of a query distribution $\beta$ (with respect to the alphabet $\Omega_n = \{X,Y,Z\}^{\otimes n}$) from which we can sample to produce our measurement bases, and other methods that involve a routine that allows us to directly produce samples of measurement bases.

To guide our discussion on randomized measurements, we introduce decision diagrams as an efficient data structure for representing query distributions and describe multiple measurement methods. As we introduce these different measurement methods, we will also mention their compatibility with the different estimators presented in Section~\ref{sec:estimators}.

\subsection{Randomized Measurements}
\begin{figure}[ht!]
    \centering
    \includegraphics[width=0.75\textwidth]{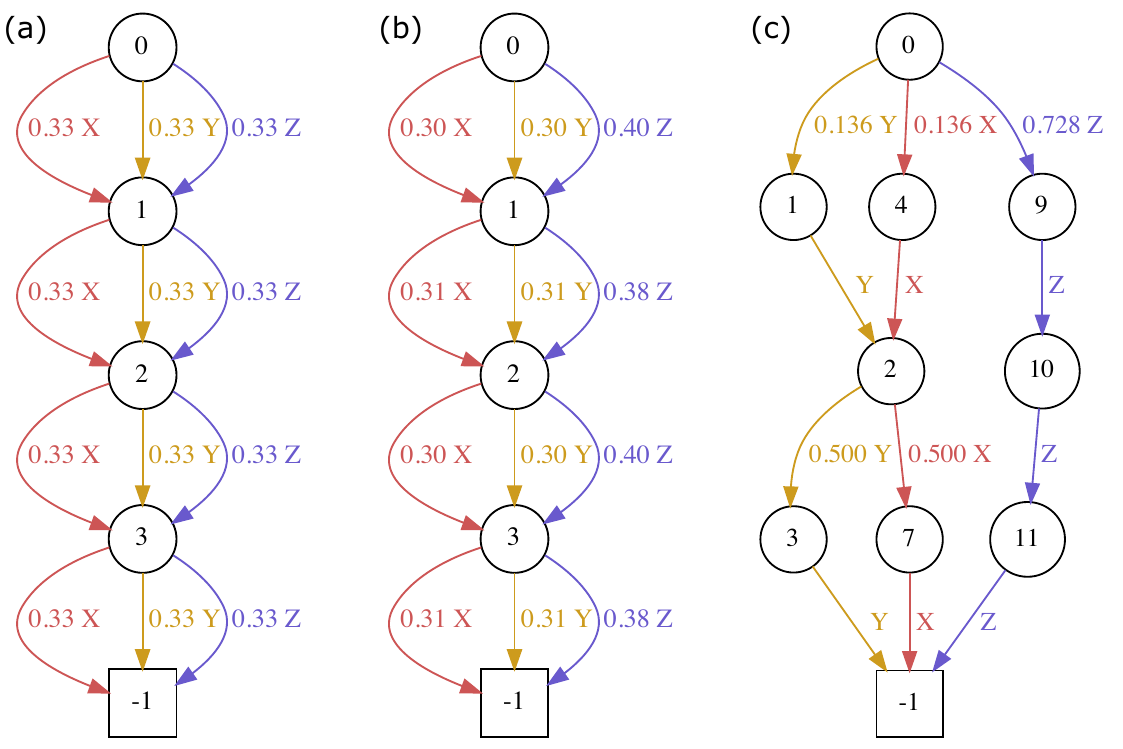}
    \caption{Instances of different decision diagrams for $H_2$ ($4$ qubits, sto3g basis, JW encoding). We show (a) uniform classical shadows, (b) locally biased classical shadows (LBCS) and (c) an optimized compact decision diagram.} \label{fig:viz_decision_diagrams_H2}
\end{figure}

% Why do we need something like decision diagrams?
We will start off by considering randomized measurement methods where the central object is a query distribution $\beta$ (with respect to the alphabet $\Omega = \{X,Y,Z\}^{\otimes n}$) from which Pauli measurement bases are sampled. 
We could describe a general $\beta$ through the probability assignments over $3^n-1$ Pauli matrices in $\Omega$. \footnote{The unassigned probability can be determined through the normalization $\sum_{x \in \Omega}\beta(x) = 1$.} However, this is exponentially expensive. Moreover, molecular Hamiltonians have $O(n^4)$ Pauli terms~\cite{mcardle2020quantum}, and these are the focus of this chapter. We could thus give probability assignments for $\beta$ over Pauli operators in $\Omega$ that cover the different Pauli terms in $H$. In fact, we can do better as many of the Pauli terms in $H$ can be grouped together, i.e.,  covered by a single Pauli measurement basis $B$.

% Introduction of decision diagrams and description
Decision diagrams~\cite{hillmich2021decision} give us a compact way to represent a query distribution $\beta$ that takes advantage of this highlighted structure in $H$ and represent probability assignments for different $B$ in $\Omega_n$ efficiently. For our purposes, decision diagrams are acyclic directed graphs. In Figure~\ref{fig:viz_decision_diagrams_H2}, we show different decision diagrams for an \ce{H_2} Hamiltonian, corresponding to different measurement methods (to be discussed later). Common to all these decision diagrams are that the parent node is set to be $0$ and the last child node is denoted by $-1$. Going from node $0$ to $-1$, we have to take a path that takes $n$ directed edges.

% How do you read a decision diagram? (or what are the different components in a dd?) Use slides.
To generate a measurement basis from a decision diagram, we follow a path from node $0$ to node $-1$, at each step choosing the next edge from the current node with probability given by its weight.
Each edge is also labeled $X$, $Y$, or $Z$, so each choice of an edge in the path represents a choice of local Pauli basis for a corresponding qubit.
Each qubit corresponds to a layer in the decision diagram, so the path from node $0$ to node $-1$ defines a choice of local basis for each layer and thus each qubit.
% To aid us reading the decision diagram, it is helpful to denote all the edges taken in the $k$th step as belonging to the $k$th layer.
% Visually in Figure~\ref{fig:viz_decision_diagrams_H2}, all the edges in the same layer are aligned.
% Each edge in any layer $k$ has a label containing an edge weight and a single qubit Pauli $P \in \{X,Y,Z\}$.
% This corresponds to the decision $B_j = P$.
% It is then easy to see that each node describes the current state e.g., node $10$ in Figure~\ref{fig:viz_decision_diagrams_H2}(c) describes the current state $B_1 B_2 = ZZ$. The edge weights are then conditional probabilities e.g., the edge from node $2$ to node $3$ has an edge weight of $P(B_3=Y|B_1 B_2 = YY) = 0.50$.
Thus decision diagrams can represent a very flexible family of query distributions $\beta$ in $O(\text{poly}(n))$ memory and allow us to efficiently generate samples from $\beta$ as well.
We now describe how different randomized measurements that have been proposed can be seen as instances of decision diagrams.

\paragraph{Classical shadows.}
The uniform classical shadows (CS) of \cite{huang2020predicting} considers the query space $\mathcal{Q} = \Omega_n = \{X,Y,Z\}^{\otimes n}$ and the query distribution $\beta(B) = 3^{-n}, \forall B \in \mathcal{Q}$. This corresponds to uniformly randomly picking a measurement basis from $\{X,Y,Z\}$ for each qubit. Despite its simplicity, it was shown in~\cite{huang2020predicting}  that for a set of $L$ Pauli observables $Q^{(j)}$, using $M = O(\log L/\epsilon^2)$ (factors depending on weight of $Q^{(j)}$ hidden) samples generated from this query distribution suffice to obtain estimates of the expectation values $\Tr(\rho Q^{(j)})$ of all $Q^{(j)}$ up to additive error $\epsilon$. It was also shown that this is asymptotically optimal.
A challenge with using CS in practice is the potentially many different circuits required to be compiled and run, while current hardware favors repetitions of the same circuit.
Moreover, CS does not incorporate any available prior information of the Hamiltonian $H$ or state $\rho$. This led to the extension of CS to other randomized methods which we will discuss shortly. We note that all the estimators (MC estimator, importance MC estimator and Bayesian estimator) discussed in Section~\ref{sec:estimators} are compatible with this measurement method.

\paragraph{Locally Biased Classical Shadows (LBCS).}
LBCS~\cite{hadfield2020measurements} is a randomized measurement method that incorporates prior information about the state $\rho$.
In LBCS, the query distribution $\beta$ over $\mathcal{Q} = \{X,Y,Z\}^{\otimes n}$ is always a product distribution
\begin{equation}
\label{lbcs_beta}
    \beta=\prod_{i=1}^n \beta_i
\end{equation}
where $\beta_i$ is the marginal probability of the $i$th qubit Pauli matrix $B_i$, but unlike CS, the $\beta_i$ are not required to be uniform.
Instead, the query distribution $\beta$ is optimized by minimizing the one-shot variance of the estimate $\hat{E}$ with respect to a reference state, subject to the constraint that it has the form \eqref{lbcs_beta}. Ideally, the reference state would be the target state, e.g., a ground state, but $\hat{E}$ is unknown \emph{a priori} for that state. Next best would be to choose a heuristic reference state in the same way one would choose an initial state for VQE or other ground state preparation algorithms, e.g., the Hartree-Fock state, but this leads to a non-convex optimization. LBCS overcomes this by instead minimizing the variance of the estimate against the maximally mixed state, which results in a convex optimization~\cite{hadfield2020measurements}. It has been numerically shown that LBCS can yield more accurate estimates of $\Tr(\rho H)$ than CS on various chemistry Hamiltonians for the same measurement budget~\cite{hadfield2020measurements}. Like CS, all the estimators (MC estimator, importance MC estimator, and Bayesian estimator) presented in Section~\ref{sec:estimators} are compatible with this measurement method.

\paragraph{Compact Decision Diagrams.}
Both of the above measurement methods are restricted to $\beta$ being a product distribution, i.e., the marginal distributions over each qubit are independent of each other.
However, in general, there will be correlations between measurements of $\Tr(\rho P_j)$ and $\Tr(\rho P_k)$, so is desirable to allow the query distribution account for such correlations.
This is where general decision diagrams fit in by allowing us to represent general non-product joint distributions efficiently for molecular Hamiltonians. We refer the reader to~\cite{hillmich2021decision,matsuo2024optderand} on how to construct compact decision diagrams given the Pauli decomposition of Hamiltonian $H$ (Eq.~\ref{eq:observable_defn}) and a qubit ordering.

% Optimization problem to be solved
Once an initial decision diagram is obtained, the edge weights of the DD can be further optimized by minimizing the one-shot variance of the estimate $\hat{E}$ if we have some prior knowledge on the quantum state $\rho$. In practice, we find that even optimizing against the maximally mixed state is beneficial, like in LBCS.
Unlike in LBCS, however, this does not yield a convex optimization problem. A solution at the local minima is still advantageous, yielding higher accuracy for the same measurement budget compared to LBCS as was shown in \cite{hillmich2021decision}.

We have so far seen how decision diagrams are useful in representing query distributions $\beta$ and how one can efficiently generate samples from $\beta$ on them. In the following sections, we discuss Adaptive Pauli Shadows (APS) and how derandomization of decision diagrams may be performed.

\subsection{Adaptive Pauli Measurements}
We have so far focused on randomized measurement methods that involve construction of a query distribution $\beta$ that we can directly sample from to generate the samples. Moreover, $\beta$ remains unchanged during the sampling process. In contrast, Adaptive Pauli Shadows (APS) method~\cite{hadfield2021adaptive} allows us to sample from $\beta$ that changes adaptively during the sampling process.

As in Algorithm~\ref{algo:energy_estimation}, we index samples by $m$.
Each time we generate a measurement basis $B^{(m)}$, we iteratively sample each single-qubit basis $B^{(m)}_{j} \in \{X,Y,Z\}$ in some qubit ordering, which is chosen randomly.
On the $j$th qubit in this ordering, we sample a measurement Pauli matrix according to the probability distribution that is the solution to 
\begin{align}
\label{aps_minimization}
    \text{minimize: } & \sum_{Q^{(k)} \in \Omega} \frac{\alpha_k^2}{\beta(Q^{(k)}_{j})} \\ \nonumber
    \text{subject to: } & 0 \leq \beta(B_{j}) \leq 1, \quad \text{for }  B_{j} \in \{X,Y,Z\} \\ \nonumber
    & \sum \limits_{B_{j} \in \{X,Y,Z\}} \beta(B_{j}) = 1
\end{align}
where the set $\Omega$ is defined as
\begin{equation}
    \Omega = \{Q \in \mathbf{Q}| Q_{j} \in \{X,Y,Z\} \text{ and } Q_{j'} \in \{I, B^{(s)}_{j'}\}~\forall j' < j \}
    \label{eq:def_set_aps}
\end{equation}
and we have used $\beta$ in place of $\beta_{j}$. An analytical solution based on Lagrange multipliers can be developed for this convex optimization problem. We point the reader to \cite{hadfield2021adaptive} for the solution. Iterating over the qubits allows us to generate the sample $B^{(s)}$. The additional cost of this sampling process is only $O(n)$.

The idea of using a distribution solving \eqref{aps_minimization} is to appropriately include the information about choices of basis on the previous qubits (in the ordering) in the choice of basis on the $j$th qubit.
The key point to notice is the definition of the set $\Omega$ in \eqref{eq:def_set_aps}: this contains all Pauli operators that are (i) in the target observable (i.e., in the set $\mathbf{Q}$), (ii) $X$, $Y$, or $Z$ on the qubit currently being decided, and (iii) compatible with the part of the basis chosen so far.

Constraints (ii) and (iii) are the important ones, and what really separate APS from other measurement methods.
Constraint (ii) is used so that the choice on qubit $j$ is independent of operators that are $I$ on qubit $j$, since these are compatible with any choice on qubit $j$. Constraint (iii) is used so that the choice on qubit $j$ is independent of operators that are already incompatible with the basis given the choices on the previous qubits.
Thus the choice of basis for qubit $j$ only depends on operators whose inclusion in the covered set actually depends on that choice.

As we do not have access to the joint query distribution $\beta$ for each sample, the Weighted-MC estimator (Eq~\ref{eq:weighted_mc_estimator}) cannot be used with this method. The MC and Bayesian estimators can be used.

\subsection{Derandomization}\label{sec:derand}
% Why do we need the concept of derandomization?
We have so far discussed randomized measurement methods that involve random generation of measurement bases. Often, however, it is desirable to have a deterministic sequence of measurement bases to make and that can be repeated across different experiments while retaining the performance of randomized algorithms.

% What do we do here?
In this section, we introduce the idea of derandomizing the randomized measurements that are obtained by sampling different query distributions $\beta$ that may correspond to CS, LBCS or in general a decision diagram. Derandomization of CS was proposed in \cite{huang2021efficient} which we extend in a straight-forward fashion to derandomization of a general query distribution $\beta$ represented on a decision diagram. Notably, we show how relevant computations carried out during derandomizing a general query distribution $\beta$ can be implemented efficiently when one has access to the corresponding decision diagram.

% Description of derandomization
Let us now introduce some notation that will be relevant to our discussion on derandomization. Recall that the target observables that we are interested in are $\mathbf{Q} = \{Q^{(j)}\}_{j \in [L]}$. Departing slightly from the goal we have considered so far, consider the goal of estimating $\Tr(\rho Q^{j})$ within error $\epsilon$ for any $j \in [L]$. We denote these estimates as $\hat{\omega}^{(j)}$ and denote the true value of $\Tr(\rho Q^{(j)})$ by $\omega^{(j)}$. This is once again achieved through $M$ single shot measurements of the following Pauli operators: $\mathbf{B} = \{B^{(s)}\}_{s \in [M]}$.

% Key idea in derandomization
Through derandomization of a randomized algorithm, it is possible to come up with a partially or fully fixed sequence of measurements to be carried out. The key idea is to understand how estimates typically deviate from the truth in a randomized algorithm and how this changes when conditioned upon previous measurements. One way to measure this deviation is through the \textit{confidence bound} introduced in \cite{huang2021efficient}:
\begin{equation}
    \mathrm{CONF}_\epsilon(\mathbf{Q};\mathbf{B}) := \sum \limits_{j=1}^L \exp \left(-\frac{\epsilon^2}{2} h(Q^{(j)};\mathbf{B}) \right),
\end{equation}
where $h(\cdot)$ is the hit function (Eq.~\ref{eq:hit_function_basis}). It was shown in \cite[Lemma 1]{huang2021efficient} that if the confidence bound is upper bounded by $\delta/2$ for some $\delta \in (0,1)$ then each of the empirical estimates $\hat{\omega}^{(j)}$ are within $\epsilon$-distance of the truth $\omega^{(j)}$, with probability at least $1-\delta$.

% What are the ingredients of derandomization?
Derandomization of DD is then completed through the following steps: (i) obtain a confidence bound on estimates for DD, (ii) analyze the confidence bound when conditioned on prior measurements, and (iii) use this conditional expectation bound to design a cost function that will be used for the derandomizing procedure. This procedure is outlined in Appendix~\ref{app_sec:derand_DD}. Here, we state the cost function in derandomization of DD.

To motivate the cost function for derandomization of DD, consider the following scenario. Suppose we are given a measurement budget of $M$ and that $\mathbf{B}^{\#}$ contains the assignments of measurement bases for the first $(m-1)$ samples and first $k$ qubits of the $m$th measurement basis. That is, we have already generated the first $(m-1)$ samples and Paulis of the first $k$ qubits of the $m$th measurement basis. We then have the following conditional expectation of the confidence bound (see Appendix~\ref{app_sec:derand_DD})
\begin{align}
    \mathbb{E}\left[ \mathrm{CONF}_\epsilon(\mathbf{Q};\mathbf{B}) | \mathbf{B}^{\#} \right] 
    =& \sum_{j \in [L]} \prod \limits_{m'=1}^{m-1} \left(1 - \eta \mathds{1}\{Q^{(j)} \triangleright {B^\#}^{(m')}\} \right) \nonumber \\
    &\times \left(1 - \eta \prod \limits_{k'=1}^{k} \mathds{1}\{Q^{(j)} \triangleright {B^\#}^{(m')}\} \Pr \left[Q^{(j)}_{k+1:n} \text{ covered by DD}| {B^{\#}}^{(m)}_{1:k} \right] \right) \nonumber \\
    & \times \left(1 - \eta \Pr \left[Q^{(j)} \text{ covered by DD}\right]  \right)^{M-m} \nonumber,
\end{align}
where $\eta = 1 - \exp(-\epsilon^2/2)$. To choose the assignment of the $k$th qubit of the $m$th measurement basis, we consider the following cost function
\begin{equation}
    {B_k^\#}^{(m)} = \argmin_{ W \in \{X,Y,Z\} } C(W) = \argmin_{ W \in \{X,Y,Z\} } \mathbb{E}\left[ \mathrm{CONF}_\epsilon(\mathbf{Q};\mathbf{B}) | \mathbf{B}^{\#}, B^{(m)}_k = W\right]
\end{equation}
where $\mathbf{B}^{\#}$ now corresponds to the assignments of measurement bases over the first $(m-1)$ samples and $(k-1)$ qubits of the $m$th measurement basis. Note that the above cost function requires the input of the experimental budget $M$.

An algorithm for derandomization of DD is given in Algorithm~\ref{algo:derandomization_DD}. Note that all the steps in the algorithm can be computed efficiently on a decision diagram for molecular Hamiltonians. Details are given in Appendix~\ref{app_sec:derand_DD}.

\begin{algorithm}[H]
    \caption{Derandomization of Decision Diagrams} 
    \textbf{Input}: Measurement budget $M$, accuracy $\epsilon$, target observables $\mathbf{Q} = \{Q^{(j)}\}_{j \in [L]}$, decision diagram $\Sigma_{DD}$ \\
    \textbf{Output}: Set of $M$ measurement bases $\mathbf{B}^{\#}$ \\
    \begin{algorithmic}[1]
        \For{$m=1:M$}
            \For{$k=1:n$}
                \For{$W \in \{X,Y,Z\}$}
                    \State Compute $C(W) = \mathbb{E}\left[ \mathrm{CONF}_\epsilon(\mathbf{Q};\mathbf{B}) | \mathbf{B}^{\#}, B^{(m)}_k = W\right]$
                \EndFor
                \State Set ${B_k^\#}^{(m)} \leftarrow \argmin_{ W \in \{X,Y,Z\} } C(W)$
            \EndFor
        \EndFor
    \State \Return $\mathbf{B}^{\#}$
    \end{algorithmic}
    \label{algo:derandomization_DD}
\end{algorithm}

The derandomized measurement procedure is compatible only with the MC estimator (Section~\ref{subsec:mc_estimator}) and Bayesian estimator (Section~\ref{subsec:bayesian_estimator}) that we have described so far. 

\paragraph*{Remark.} Both measurement methods of APS and derandomization attempt to bring in adaptivity into the sampling process. However, there are qualitative differences besides how these methods themselves are motivated and set up. Derandomization fixes the Pauli operator for a qubit by solving an optimization problem while APS obtains an optimized marginal distribution over Paulis on a qubit and then allows for a randomized Pauli sample. Measurement history is taken into account of derandomization but not in APS.

\section{\benchname:~\benchabbr}\label{sec:CHSCQCBench}
Having discussed different randomized and derandomized measurement methods at our disposal for the learning problem of estimating $\Tr(\rho H)$ (Section~\ref{sec:background}) in Section~\ref{sec:meas_methods} and estimators that can be used in conjunction with these methods in Section~\ref{sec:estimators}, we are now in a position to compare the performance of these different measurements in practice. 

In Section~\ref{subsec:CHSCQCBench_strategy}, we laid the inspiration and strategy for \benchname. We now formally describe the setup, execution and following analysis, discussed in the context of general measurement protocols in Section~\ref{subsec:general_protocol_meas}, to be carried out as part of this benchmark. \benchname~along with the implementation of all measurement methods (Section~\ref{sec:meas_methods}) and estimators (Section~\ref{sec:estimators}) can be found in a Github repository \footnote{\url{https://github.com/arkopaldutt/RandMeas}}.

\subsection{Common quantum computation task and common objects}
In Section~\ref{subsec:CHSCQCBench_strategy}, we mentioned the types of Hamiltonians and states that we would benchmark against to assess the performance of different measurement methods. We now explicitly state these objects. For \benchname, we consider a set of small molecular electronic Hamiltonians that have been encoded into qubit systems. The molecules that we specifically consider as part of the benchmark here, are listed in Table~\ref{table:molecules}. The qubit Hamiltonians are obtained by first representing each molecule by a fermionic Hamiltonian in a particular molecular orbital basis which is then mapped to a qubit Hamiltonian using the Jordan-Wigner (JW) \cite{jordan1993paulische} encoding. In our experiments we found similar results against the Bravyi-Kitaev \cite{bravyi2002fermionic}, and parity \cite{bravyi2002fermionic,seeley2012bravyi} encodings. We thus only report our results for the JW mapping.

As mentioned earlier in Section~\ref{subsec:CHSCQCBench_strategy}, given a Hamiltonian $H$, the set of states considered are those which we expect during the runtime of a hybrid quantum-classical circuit for ground state estimation such as VQE. These include a properly chosen initialization (e.g., Hartree-Fock (HF) state), target state (e.g., ground state) and appropriately chosen parametrized quantum circuits (also called quantum ansatz). For the purpose of the benchmarking experiments in this paper, we consider the HF state, ground state and a random quantum ansatz. Another motivation for including random quantum ansatz as one of the states to test on is simply because in most cases, the ground state is difficult to prepare on a near-term quantum device and it is the state that the hybrid quantum-classical algorithm is trying to optimize for.

An overview of the different molecules and the experiments carried out on them are given in Table~\ref{table:molecules}. These include tapered Hamiltonians~\cite{bravyi2017tapering} where qubit degrees of freedom corresponding to exact $Z_2$ symmetries of a Hamiltonian are removed. At least two qubits can always be removed, corresponding to spin and electron number parity, and often more $Z_2$ symmetries are available, which can typically be attributed to the point group of the molecule~\cite{bravyi2017tapering,setia2020pointgroup}. 

We will describe the experimental protocol followed in the execution of \benchname~in Section~\ref{subsec:expt_setup}. It should be noted that while we consider a small set of Hamiltonians in this work, \benchname~can be performed on any set of molecular Hamiltonians such as those included in the QM databases~\cite{blum2009qmdata,ruddigkeit2012qm9data}.

\begin{table}[ht!]
\small
\centering
\begin{tabular}{|l|l|l|}
\hline
\textbf{Molecule} & \textbf{Number of qubits} & \textbf{Basis}  \\ \hline
$\mathrm{H_{2}}$ (tapered)   & 5  & 3-21g \\  
\hline
$\mathrm{HeH^{+}}$ (tapered)   & 6  & 3-21g \\ 
\hline
$\mathrm{HeH^{+}}$   & 8    & 6-31g \\ 
\hline
$\mathrm{LiH}$      & 12    & sto6g \\ 
\hline
$\mathrm{N_2}$     & 16     & sto6g \\ 
\hline
%  HF (dev.)
% $\mathrm{H_2}$      & 20    & CCPVDZ & JW & random ansatz (dev.) \\ 
% \hline
% $\mathrm{LiH}$      & 22    & 6-31g & JW & random ansatz (dev.) \\ 
% \hline
\end{tabular}
\caption{Molecules considered for benchmarking of measurement methods on the simulator and quantum device. The Jordan Wigner encoding is considered. For each Hamiltonian, the state is set to be the ground state on the simulator and a random ansatz on quantum devices.}
\label{table:molecules}
\end{table}
\vspace{-2em}

\subsection{Performance metrics in \benchname}\label{sec:performance_metrics_benchmark}
In this section, we describe the different metrics considered as part of the analysis stage of benchmarking different measurement methods in \benchname. In Figure~\ref{fig:pep_sea_viz}, we showcased two metrics of relevance, accuracy and runtime, which were later contextualized in Section~\ref{subsec:general_protocol_meas}.

\paragraph{Accuracy.} In assessing measurement methods, the main performance metric is the accuracy or learning error achieved given a measurement budget $M$. For various measurement protocols, we track the root mean squared error (RMSE) with increasing values of measurement budget. This also allows us to investigate the convergence behaviour of different measurement methods and numerically determine their sample complexities.

\paragraph{Runtimes.} As we identified in Section~\ref{subsec:general_protocol_meas} and illustrated in Figure~\ref{fig:pep_sea_viz}, each step of the measurement protocol (pre-processing, experiment, and post-processing) has a computational runtime associated with it. For the pre-processing and post-processing steps, this is the classical computational runtimes as these are executed on classical computing. For experiments, this is a classical computational runtime when executed on a simulator and the quantum wall-clock time when executed on a quantum device. Here, we keep track of classical computational runtime as wall-clock time instead of number of FLOPS (floating point operations). It should be noted that these computational runtimes are then specific to our implementation of the measurement methods and estimators.

\paragraph{Classical latencies.}
To execute experiments on a quantum computer according to the measurement bases generated on a classical computer, there are multiple latencies involved such as compilation of circuits to the native gate set of the quantum computer, loading of circuits through the control electronics interfacing with the quantum computer, and classical post-processing associated with measurement error mitigation. On current control electronics, it is more expensive to run any number of shots over many different circuits than the same number of shots against the same circuit. Thus, we also track the number of unique measurement circuits (or measurement bases) requested by different measurement methods to reach a given value of accuracy and the distribution of shots across different circuits.

\paragraph{Summary of resource utilization.}
% HHL algorithm, Accessibility
% Rate of logical instructions
As suggested in Section~\ref{subsec:CHSCQCBench_metrics_design}, we attempt to summarize resource requirements by answering the following question: \emph{How much classical and quantum resources are utilized by a given measurement method $\Sigma_M$ to reach a cutoff of accuracy $5$ milli Hartree in estimating $\Tr(\rho H)$ for a given Hamiltonian $H$ and $\rho$?}

Ideally, we would have chosen the cutoff to be chemical accuracy or $1.593$ milli Hartree. However, many of the candidate measurement methods considered here would require far too high measurement budget to reach this cutoff and would not be possible to verify this reasonably in experiment on quantum devices. In the above question, we have chosen $5$ milli Hartree as a cutoff as this is achieved by various measurement methods on a simulator and in experiments on a quantum device. The resource utilization at this cutoff should be representative of that at chemical accuracy. In answering the above question, we will primarily account for classical and quantum runtimes. We will ignore classical latencies as control electronics hardware is fast evolving.

As noted earlier, we cannot simply add the classical and quantum wall-clock times to obtain a value associated with overall resource utilization due to the differing maturities of classical and quantum computing technologies. Rather, we introduce a heuristic for resource utilization that takes the form of a weighted sum of the wall-clock times:
\begin{equation}
    \mathrm{R} = w_c \cdot R_{\mathrm{classical}} + w_q \cdot R_{\mathrm{quantum}},
    \label{eq:resource_utilization_heuristic}
\end{equation}
where $w_c$ is the weight corresponding to classical computers, $w_q$ is the weight corresponding to quantum computers and $R_{\mathrm{classical}}$ (or $R_{\mathrm{quantum}}$) is the resource utilization by the corresponding type of computing and measured in wall-clock time here. The weights have units of $\text{s}^{-1}$ to make $\mathrm{R}$ dimensionless and can be interpreted as resources used per second.

The question then arises: How do we design these weights? We could consider a common task for both classical and quantum computers, and then compare their performances. For example, in comparing CPU-centric classical computers and those with access to GPUs, a common task is solving pseudorandom dense linear systems. Towards this, LINPACK has been extended for these computers~\cite{dongarra2003linpack,jo2015multinodes,petitet2018hpl}. Here, as well, we could consider the common task of solving random linear systems of equations on both classical and quantum computers, albeit with different input and output models. However, there have not been any experiments on quantum devices solving large-scale linear systems using the HHL algorithm. 

Another way to design these weights is to consider the rate of logical instructions. A natural metric for classical computers is that of FLOPS (floating point operations per second) and for quantum computers is that of CLOPS (circuit layer operations per second)~\cite{wack2021scale}. A general purpose classical computer has a speed of around $6$ GFLOPS and the IBM quantum computer (\texttt{ibmq\_mumbai}) which we primarily used for our experiments had a speed of around $2.4$ KCLOPS during our experiments. Comparing these speeds gives us weights of ($w_c$,~$w_q$)~=~($1$,~$2.5 \times 10^6$). However, designing weights in such a manner has flaws of neglecting the finite coherence time of quantum computers, and not accounting for different energy costs of these computing resources. To get around this, we instead suggest different regimes for weights in Table~\ref{table:weights_heuristic} and which allows the user of the benchmark to incorporate their own preferences.

\begin{table}[ht!]
\centering
\small
{\renewcommand{\arraystretch}{1.2}
\begin{tabular}{|l|r|}
\hline
\textbf{Regime}    & \multicolumn{1}{l|}{\textbf{Weights ($w_c$, $w_q$)}} \\ \hline
A: Fast QPU        & ($1$, $1$) \\ \hline
B: Medium Fast QPU & ($1$, $1.5 \times 10^2$) \\ \hline
C: Medium Slow QPU & ($1$, $2 \times 10^4$)   \\ \hline
D: Slow QPU        & ($1$, $2.5 \times 10^6$) \\ \hline
\end{tabular}
}
\caption{Different regimes of weights for classical and quantum computers in our heuristic.}
\label{table:weights_heuristic}
\end{table}
\vspace{-1em}

\subsection{Experimental protocol for \benchname} \label{subsec:expt_setup}
% What are the methods that we benchmark against currently? In terms of measurement methods, estimators
We have so far described the objects in \benchname~and the different performance metrics considered as part of the analysis in \benchname. Let us cross the rest of our SEAs. We will benchmark against the following candidate measurement methods, previously described in Section~\ref{sec:meas_methods}, and included as part of the pre-processing setup: uniform classical shadows (CS), locally biased classical shadows (LBCS), decision diagrams (DD), and derandomization of these methods (Derand. CS, Derand. LBCS and Derand. DD) respectively. We do not include APS (see Appendix~\ref{app_sec:convergence_APS}). In addition to these measurement methods, we will carry out experiments with different estimators (Section~\ref{sec:estimators}). 

We will now describe the experimental protocol followed for each combination of measurement procedure and estimator on a given Hamiltonian. Experiments are executed on an ideal classical simulator and IBM quantum devices. On the simulator, we assume that copies of an unknown quantum state $\rho$ are given to us. On the quantum device, we assume access to a quantum circuit preparing the unknown quantum state $\rho$. This may be in the form of an ansatz as one would have during a step of the VQE algorithm. In each experiment, the goal is to then estimate the energy $E=\Tr(\rho H)$ of the encoded $n$-qubit Hamiltonian $H$. 

To compare the performance of the different measurement procedures, we use the metric of root mean square error (RMSE) of the ground state energy against a budget of $M$ Pauli measurements. The measurement budget $M$ will be specified for different cases later. We compute RMSE by independently repeating each experiment with measurement budget $M$, $N$ times so that we have access to $N$ independent estimates of the energy $\{\hat{E}^{(i)}_G\}_{i \in [N]}$. RMSE is then computed as
\begin{equation}
    \mathrm{RMSE} = \sqrt{ \frac{1}{N} \sum \limits_{i=1}^N \left( \hat{E}^{(i)}_G - E_G\right)^2},
    \label{eq:rmse_definition}
\end{equation}
where $E_G$ is the true energy. We have access to the true energy on the simulator. On the devices, we replace this with the empirical mean computed over $N$ runs.

%State experimental constraints  and computational constraints such as 4 days on supercloud
\paragraph{Classical simulator} The ground state and ground energy of each $n$-qubit molecular Hamiltonian is determined through the Lanczos method \cite{martinsson2020randomized}. The simulator is then initialized with this ground state and this quantum circuit is provided to the measurement procedure in lieu of $\rho$. As described in Section~\ref{subsec:general_protocol_meas}, depending on the Pauli measurement basis, this quantum circuit is then appended with a measurement circuit and the qubits are measured in computational basis to produce a single shot (of measurement outcomes). All the simulator runs are executed on a cluster which has $\texttt{Intel Xeon Platinum 8260 (2.4 GHz)}$ nodes. The classical computational runtimes reported here correspond to these runs and are specific to this CPU. A constraint from the cluster is that all simulations must be completed within four days of wall-clock time and this is imposed on our runs as well. This sets a constraint for the measurement budget on some of the measurement methods being considered.

\paragraph{Quantum device} We also benchmark the different measurement methods on IBM quantum devices which include the 16-qubit \texttt{ibmq\_guadalupe} and 27-qubit \texttt{ibmq\_mumbai}. Rather than considering access to the ground state as on the simulator, we consider the state preparation circuit to be an excitation preserving ansatz \cite{barkoutsos2018quantum} of depth $25$ for our experiments. We then compare the performance of different measurement procedures on the ansatz with a set of (fixed) randomly assigned parameters. A relevant constraint from the hardware side is that each job on the IBM quantum devices are limited to have $300$ different circuits and $10^5$ shots each. Further, a sequence of jobs is given a maximum time allocation of $8$ hours. As we will see later, this constrains different measurement methods (e.g., CS and LBCS) which generate many unique circuits, that can be benchmarked on the quantum device on large molecules.

\section{Results}\label{sec:results}
In this section, we carry out a systematic comparison of the performance of different measurement methods (Section~\ref{sec:meas_methods}) in \benchname~(Section~\ref{sec:CHSCQCBench}). Previous comparisons have largely focused on analytical single shot variances against different molecules~\cite{hadfield2020measurements,hillmich2021decision} or small fixed measurement budgets~\cite{huang2021efficient,hadfield2021adaptive,shlosberg2023adaptiveestimation}. Moreover, measurement methods are often equipped with different estimators, making it less clear how much gain in performance one obtains by switching measurement methods.

We first pay special attention to the convergence of accuracy of these measurements. We then evaluate different measures of classical and quantum resource utilization as discussed in Section~\ref{sec:performance_metrics_benchmark} for each measurement method. We use the experimental protocols as discussed in Section~\ref{subsec:expt_setup}. In Section~\ref{subsec:comparison_estimators}, we evaluate the RMSE of a specified measurement method when equipped with different estimators. This is to highlight the advantage one can gain in terms of the number of shots required to achieve a desired accuracy in estimation of $\Tr(\rho H)$ by simply changing the estimator. In Section~\ref{sec:ideal_sim_results}, we report results from \benchname~on a simulator and in Section~\ref{sec:device_results} on quantum devices for different molecular Hamiltonians and states~\ref{table:molecules}. Finally, we comment on the utilization of quantum and classical resources with experiments on IBM quantum devices in Section~\ref{sec:device_results}.
% Finally, we comment on further progress required to be made to evaluate energies at chemical accuracy and that of larger Hamiltonians in Section~\ref{subsec:large_expts}.

\subsection{Comparison of estimators}\label{subsec:comparison_estimators}
In Section~\ref{sec:estimators}, we noted that there are multiple different estimators~--- Monte Carlo (MC), weighted MC (WMC) and Bayesian~--- that could be used with the various measurement methods. We also noted that asymptotically with proper choice of parameters, all these estimators give the same performance in terms of the expected value of $\Tr(\rho H)$ for a given Hamiltonian $H$. This has motivated the use of different estimators combined with different measurement methods in previous comparison studies~\cite{hadfield2020measurements,huang2021efficient,wu2023overlapped}. However, this becomes problematic when the measurement budget (or the total number of shots) considered is very low as has been the case in these studies and we are not in the asymptotic regime where the estimators are equivalent. The difference in performance of two combinations of measurements methods with estimators cannot be then properly attributed to the difference in estimators or the difference in measurement methods.

\begin{figure}[H]
    \centering
    \includegraphics[width=\textwidth]{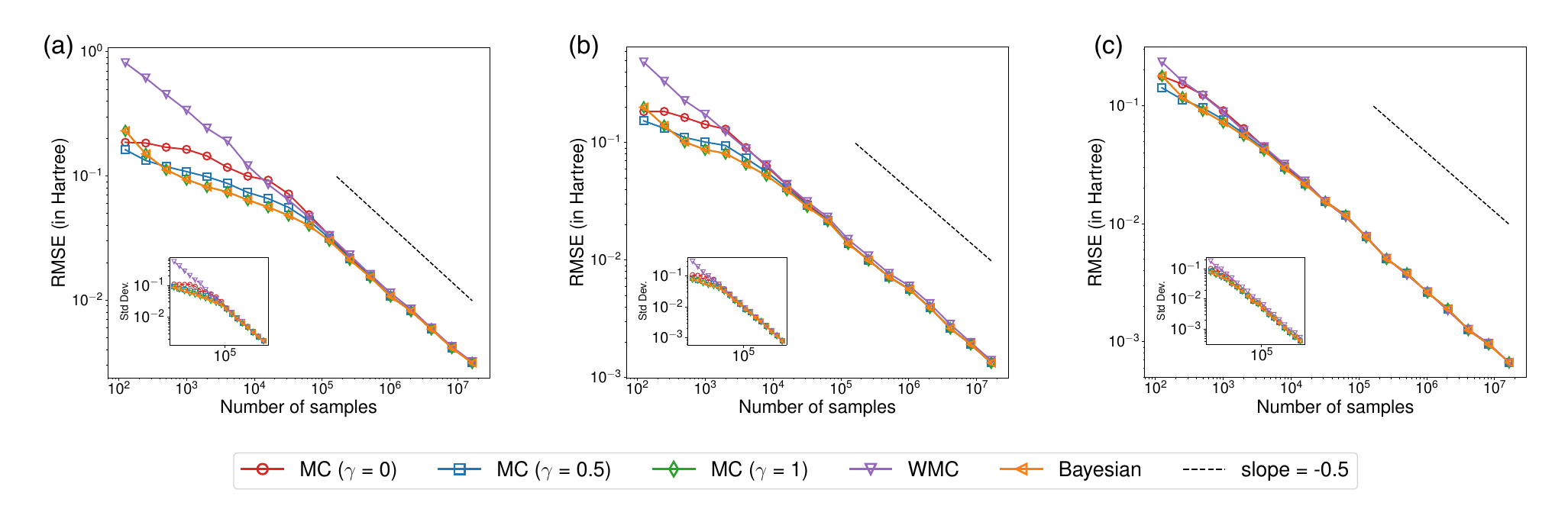}
    \caption{Comparison of RMSE achieved by different estimators in numerical simulations on $\ce{HeH^{+}}$ cation (8 qubits, 6-31g basis, JW encoding) with measurement methods of (a) classical shadows (CS), (b) locally biased classical shadows (LBCS), and decision diagrams (DD). The state is set to be the ground state of the Hamiltonian. A common legend is shown for the subfigures. Trends of RMSE achieved by Monte Carlo (MC) with Laplace smoothing of $\gamma \in \{0,0.5,1\}$, weighted Monte Carlo (WMC) and Bayesian estimators on the same sets of measurements collected for each measurement method are shown. Averaged values over $200$ independent runs are shown. Inset shows the standard deviation or uncertainty on these expectation values for each estimator.}
    \label{fig:comparison_est_HeH_sim}
\end{figure}

To highlight the advantage one can gain in terms of RMSE achieved in estimating $\Tr(\rho H)$ for a low measurement budget and how this apparent advantage disappears by increasing the number of shots, we consider the problem of estimating the value of $\Tr(\rho H)$ for a fixed measurement method and different estimators. As an illustration, we set the Hamiltonian $H$ to be that of $\ce{HeH^{+}}$ cation (6-31g basis, JW encoding, $8$ qubits) and $\rho$ as its ground state. In Figure~\ref{fig:comparison_est_HeH_sim}(a), we plot the trends of RMSE achieved by the CS method with different estimators for $\ce{HeH^{+}}$. We observe that at a measurement budget of $10^3$ shots (or samples), the Bayesian estimator has a lower RMSE than the MC estimator (although this difference can be removed by increasing the smoothing factor $\gamma$ to $1$), which in turn has a lower RMSE than the WMC estimator. These differences however disappear after $10^5$ shots. Similarly, in Figure~\ref{fig:comparison_est_HeH_sim}(b) for the LBCS measurement method, any advantage offered by one estimator disappears after around $10^4$ shots. This is also observed in Figure~\ref{fig:comparison_est_HeH_sim}(c) for the case of optimized decision diagrams (DD).

The implications of these results are twofold. Firstly, we need to be systematic in our choice of estimator when comparing different measurement methods for low measurement budgets so that we can properly attribute differences in performance to the measurement method at hand. Secondly, for low measurement budgets, a Bayesian estimator or MC estimator with Laplace smoothing is preferred. In the rest of the paper, we will either fix the estimator across all the measurement methods or state when different estimators are chosen for different measurement methods. We can afford to do the latter as there is negligible difference between any of the estimators for a given measurement method at the high measurement budgets ($>10^6$ shots) that we consider.

\vspace{-2em}

\subsection{Experiments on classical simulator} \label{sec:ideal_sim_results}
We now turn our attention to comparing the performance of different measurement methods in estimating $\Tr(\rho H)$ on a classical simulator for molecular Hamiltonians (those given in Table~\ref{table:molecules}) and their ground states $\rho$. In all the following experiments, the estimator will be set to be the Bayesian estimator with the exception of \ce{N_2} where it is set to be the MC estimator. The latter choice is made to reduce the classical post-processing runtime for \ce{N_2}.

\begin{figure}[H]
    \centering
    \includegraphics[width=0.6\textwidth]{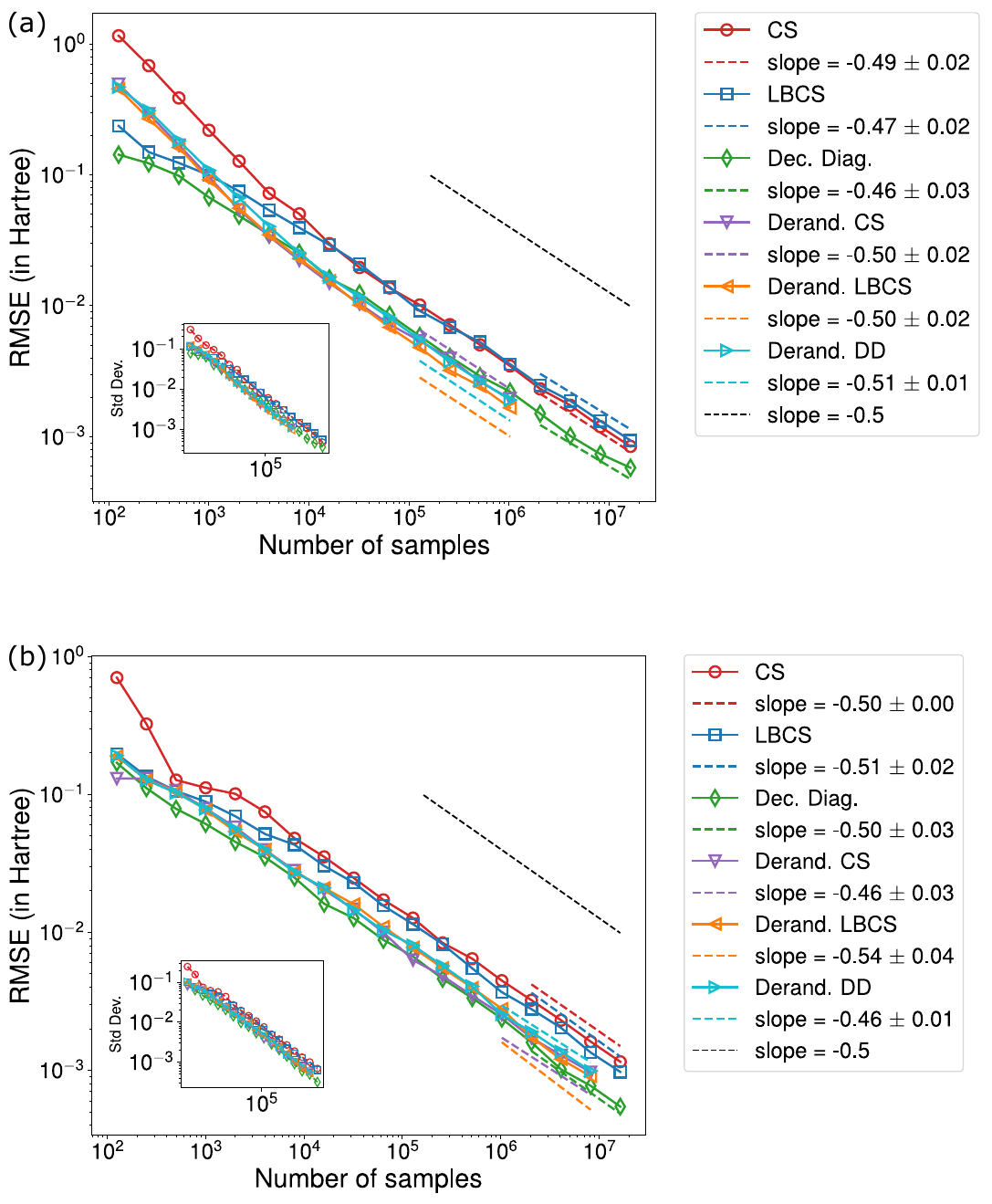}
    \caption{Comparison of RMSE achieved in numerical simulations by different measurement methods in estimating $\Tr(\rho H)$ with $\rho$ set as the ground state and $H$ is the Hamiltonian of (a) tapered \ce{H_2} ($5$ qubits, 3-21g basis, JW encoding), and (b) tapered \ce{HeH^{+}} ($6$ qubits, 3-21g basis, JW encoding). RMSE is shown with the number of samples (or shots) made. The estimator for each measurement method is set to be the Bayesian estimator.}
    \label{fig:comparison_meas_methods_H2_5q_HeH_6q_sim}
\end{figure}

\begin{figure}[H]
    \centering
    \includegraphics[width=0.6\textwidth]{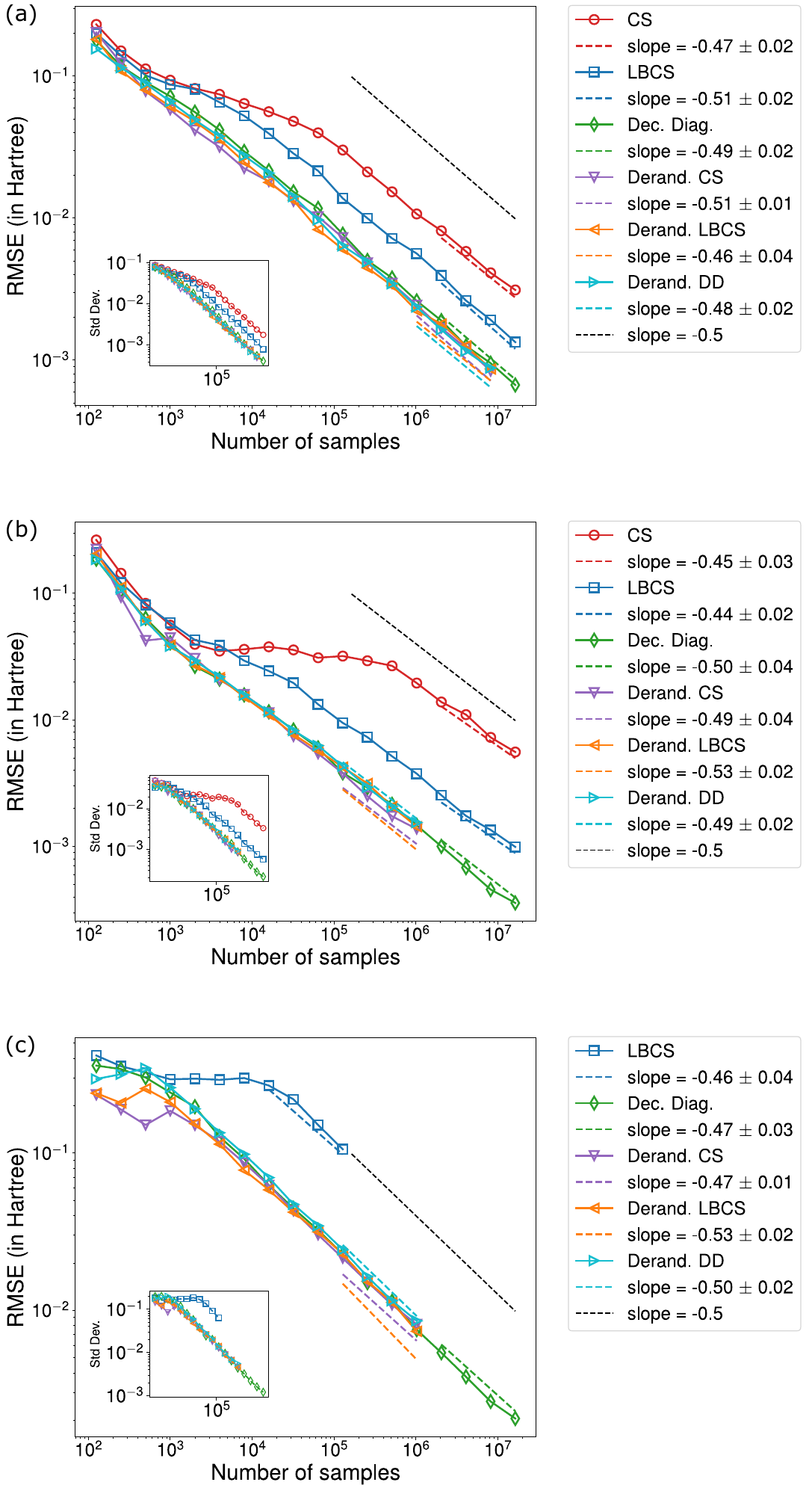}
    \caption{Comparison of RMSE achieved in numerical simulations by different measurement methods in estimating $\Tr(\rho H)$ with $\rho$ set as the ground state and $H$ is the Hamiltonian of (a) \ce{HeH^{+}} ($8$ qubits, 6-31g basis, JW encoding), (b) \ce{LiH} ($12$ qubits, sto-6g basis, JW encoding), and (c) \ce{N_2} ($16$ qubits, sto-6g basis, JW encoding). RMSE is shown with the number of samples (or shots) made. The estimator for each measurement method is set to be the Bayesian estimator in (a)-(b) and the MC estimator in (c).}
    \label{fig:comparison_meas_methods_HeH_8q_LiH_12q_N2_16q_sim}
\end{figure}

Figures~\ref{fig:comparison_meas_methods_H2_5q_HeH_6q_sim} and \ref{fig:comparison_meas_methods_HeH_8q_LiH_12q_N2_16q_sim} show convergence behaviour of RMSE achieved by different measurement methods against measurement budget up to around $1.6 \times 10^7$ samples. In Table~\ref{tab:resource_analysis_sim}, we summarize resource utilization by different measurement methods against different molecular Hamiltonians considering metrics discussed in Section~\ref{sec:CHSCQCBench}. We now describe our observations for the set of molecular Hamiltonians in Table~\ref{table:molecules}.

\paragraph{Tapered Hamiltonians.} We show convergence results for the tapered Hamiltonian \ce{H_2} (3-21g, JW, 5 qubits) in Figure~\ref{fig:comparison_meas_methods_H2_5q_HeH_6q_sim}(a) and tapered Hamiltonian \ce{HeH^{+}} (3-21g, JW, 6 qubits) in Figure~\ref{fig:comparison_meas_methods_H2_5q_HeH_6q_sim}(b). We observe that CS performs very similar to LBCS for these smaller sized Hamiltonians. Derandomization methods also perform very similarly to decision diagrams with Derand. LBCS narrowly improving on \ce{H_2} and decision diagrams on \ce{HeH^{+}}.

\paragraph{8 to 16 qubit Hamiltonians.} We show results for \ce{HeH^{+}} (6-31g, JW, $8$ qubits) in Figure~\ref{fig:comparison_meas_methods_HeH_8q_LiH_12q_N2_16q_sim}(a), \ce{LiH} (sto6g, JW, $12$ qubits) in Figure~\ref{fig:comparison_meas_methods_HeH_8q_LiH_12q_N2_16q_sim}(b), and \ce{N_2} in Figure~\ref{fig:comparison_meas_methods_HeH_8q_LiH_12q_N2_16q_sim}(c). Across all these three molecular Hamiltonians, we find that optimized decision diagrams (Dec. Diag.) and derandomization (Derand.) are best, achieving the same RMSE compared to CS or LBCS with fewer shots. In particular for \ce{HeH^{+}}, decision diagrams are able to achieve chemical accuracy of $1.6$ milli-Hartree, and require around $22\%$ fewer shots than LBCS. By using LBCS itself, we obtain a constant query reduction (as observed after $2 \times 10^4$ shots) of $40\%$ compared to CS.

Similarly, for \ce{LiH} we notice a large gap in performance in between LBCS and CS versus decision diagrams and any of the derandomized methods. Overall, in this case, Derand. CS is the preferred measurement method, shaving off nearly two orders of magnitude of shots required to reach high accuracy compared to CS, although it is extremely close to the other derandomized methods and decision diagrams. On \ce{N_2}, we do not benchmark CS given its poor performance on previous molecules and the high number of unique measurement circuits needed to be evaluated on the simulator. On \ce{N_2}, we observe that Derand. LBCS is the most accurate method followed by decision diagrams.

\paragraph{Resource utilization.}
In Table~\ref{tab:resource_analysis_sim}, we report the resource utilization for estimating $\Tr(\rho H)$ within an accuracy cutoff of $5$ milli Hartree. While we have so far commented on accuracy, let us consider the other metrics. We notice that across the board, Derand. DD requires the least number of unique circuits to be executed. This has implications for reduced time for quantum compilation of measurement circuits on quantum hardware. Regarding classical pre-processing runtime, generating samples from product distributions is fast which makes CS and LBCS attractive in this respect. For decision diagrams (DD), the largest contribution to the classical pre-processing runtime is the runtime taken to construct and optimize the decision diagrams with time taken to generate samples for $5$ milli Hartree accuracy being $10$s of seconds up to \ce{LiH} and taking roughly $490$s for \ce{N_2}. The runtimes for constructing DDs are reported in Appendix~\ref{app_sec:molecules_query_distrn}. For Derand DD, the largest contribution to classical pre-processing runtime is the derandomization process itself. In terms of total runtime on the simulator, decision diagrams and derandomization start becoming preferred over CS and LBCS as we move to the larger molecules of \ce{LiH} and \ce{N_2}.

In Table~\ref{tab:resource_analysis_sim_predicted_for_dev}, we report estimated resource utilization for estimating $\Tr(\rho H)$ within an accuracy cutoff of $5$ milli Hartree. The quantum runtimes are predicted assuming that most of the quantum wall-clock time is due to delay between executions of circuits which is around $500\mu$s on the quantum devices which we ran our experiments on. This is an example of how CSHOREBench could be used as a tool for selecting measurement methods before running any experiment on the quantum device.

\begin{table}[H]
\centering
\footnotesize
{\renewcommand{\arraystretch}{1.2}
\begin{tabular}{|l||l|r|r|l|r|r|l|}
\hline
% \begin{tabular}[c]{@{}c@{}}Molecular\\ Hamiltonian\end{tabular} & 
\multicolumn{1}{|c|}{\begin{tabular}[c]{@{}c@{}}Measurement \\ Method\end{tabular}} & 
\multicolumn{1}{c|}{\begin{tabular}[c]{@{}c@{}}\# shots\\ required\end{tabular}} & 
\multicolumn{1}{c|}{\begin{tabular}[c]{@{}c@{}}\# unique\\ circuits \end{tabular}} & 
\multicolumn{1}{c|}{\begin{tabular}[c]{@{}c@{}}Median \#\\ of shots per circuit\\ (all, top 5\%, bottom 5\%)\end{tabular}} & 
\multicolumn{1}{c|}{\begin{tabular}[c]{@{}c@{}}Classical\\ pre-process.\\ runtime [s]\end{tabular}} & 
\multicolumn{1}{c|}{\begin{tabular}[c]{@{}c@{}}Classical\\ post-process.\\ runtime [s]\end{tabular}} & 
\multicolumn{1}{c|}{\begin{tabular}[c]{@{}c@{}}Classical\\ simulator\\ runtime [s]\end{tabular}} & 
\multicolumn{1}{c|}{\begin{tabular}[c]{@{}c@{}}Total \\ classical \\ runtime [s] \end{tabular}} 
\\ \hline \hline
\multicolumn{8}{|c|}{\cellcolor[HTML]{EFEFEF}\ce{H_2}, 5 qubits (3-21g, JW)} \\ \hline
CS           & $4.75 \times 10^5$ & $243$  & (1954, 1954, 1954)     & $1.21 \times 10^2$ & $4.94$ & $4.27$  & $1.30 \times 10^2$ \\ \hline
LBCS         & $5.11 \times 10^5$ & $243$ & (944, 12350, 94)        & $9.59 \times 10^1$ & $7.13$ & $3.73$  & $1.07 \times 10^2$ \\ \hline
Dec. Diag.   & $1.79 \times 10^5$ & $44$ & (1335, 15122, 36)        & $\mathbf{4.39 \times 10^1}$ & $1.59$  & $0.91$  & $\mathbf{4.64 \times 10^1}$ \\ \hline
Derand. CS   & $1.46 \times 10^5$ & $49$ & (4921, 5628, 1)          & $1.72 \times 10^3$ & $\mathbf{1.52}$ & $0.88$  & $1.72 \times 10^3$ \\ \hline
Derand. LBCS & $\mathbf{1.16 \times 10^5}$ & $55$ & (18, 4462, 1)   & $2.31 \times 10^3$ & $1.62$ & $0.92$  & $2.31 \times 10^3$ \\ \hline
Derand. DD   & $1.53 \times 10^5$ & $\mathbf{32}$ & (5453, 5461, 1) & $1.86 \times 10^3$ & $1.67$ & $\mathbf{0.71}$  & $1.86 \times 10^3$ \\ \hline
% Dec Diag: 10^1 \times (2.88 (for opt DD) + 1.51 (sampling))
% Derand DD: 2.88e1 (for opt DD) + 1.84e3 (sampling)
\hline
\multicolumn{8}{|c|}{\cellcolor[HTML]{EFEFEF}\ce{HeH^+}, 6 qubits (6-31g, JW)} \\ \hline
CS           & $8.25 \times 10^{5}$ & $729$  & (1132, 1132, 1132) & $1.52 \times 10^2$ & $30.63$ & $13.50$  & $1.96 \times 10^2$ \\ \hline
LBCS         & $6.35 \times 10^{5}$ & $729$  & (372, 4078, 37)    & $\mathbf{1.16 \times 10^2}$ & $18.19$ & $11.40$  & $\mathbf{1.46 \times 10^2}$  \\ \hline
Dec. Diag.   & $\mathbf{2.13 \times 10^{5}}$ & $234$  & (253, 5718, 7)     & $5.63 \times 10^2$ & $\mathbf{5.60}$ & $3.41$  & $5.72 \times 10^2$ \\ \hline 
Derand. CS   & $2.37 \times 10^{5}$ & $188$  & (665, 2790, 1)     & $5.24 \times 10^3$ & $8.80$ & $2.59$  & $5.25 \times 10^3$ \\ \hline
Derand. LBCS & $3.11 \times 10^{5}$ & $184$  & (950, 3616, 1)     & $7.13 \times 10^3$ & $8.91$ & $2.64$  & $7.14 \times 10^3$ \\ \hline 
Derand. DD   & $3.06 \times 10^{5}$ & $\mathbf{123}$  & (3589,3610, 1)     & $6.08 \times 10^3$ & $8.04$ & $\mathbf{2.10}$  & $6.09 \times 10^3$ \\ \hline
% Dec Diag: 5.45e2 (for opt DD) + 1.78e1 (sampling)
% Derand DD: 5.45e2 (for opt DD) + 5.53e3 (sampling)
\hline
\multicolumn{8}{|c|}{\cellcolor[HTML]{EFEFEF}\ce{HeH^+}, 8 qubits (6-31g, JW)} \\ \hline
CS           &  $5.58 \times 10^{6}$          & $6561$& (850, 850, 850) & $1.14 \times 10^3$ & $575.43$  & $2.06 \times 10^2$  & $1.92 \times 10^3$ \\ \hline
LBCS         &  $9.96 \times 10^{5}$          & $6561$& (81, 1001, 10)  & $\mathbf{2.47 \times 10^2}$ & $105.92$  & $1.44 \times 10^2$  & $\mathbf{4.97 \times 10^2}$ \\ \hline
Dec. Diag.   &  $2.92 \times 10^{5}$          & $781$ & (99, 2907, 3)   & $3.57 \times 10^3$ & $25.33$   & $1.96 \times 10^1$  & $3.61 \times 10^3$ \\ \hline
Derand. CS   &  $2.56 \times 10^{5}$          & $545$ & (43, 2345, 1)   & $9.16 \times 10^3$ & $26.40$  & $1.30 \times 10^1$  & $9.20 \times 10^3$ \\ \hline
Derand. LBCS &  $\mathbf{2.05 \times 10^{5}}$ & $517$ & (24, 1869, 1)   & $7.52 \times 10^3$ & $21.80$  & $1.18 \times 10^1$  & $7.55 \times 10^3$ \\ \hline
Derand. DD   &  $2.34 \times 10^{5}$          & $\mathbf{209}$ & (1407, 1960, 1) & $9.94 \times 10^3$ & $\mathbf{20.30}$  & $\mathbf{7.13}$  & $9.97 \times 10^3$ \\ \hline
% For CS, LBCS, DD, we carried out reanalysis runs. Post-process times were estimated from Derand counterparts.
% Dec Diag: 3.54e3 (for opt DD) + 3.07e1 (sampling)
% Derand DD: 3.54e3 (for opt DD) + 6.40e3 (sampling)
\hline
\multicolumn{8}{|c|}{\cellcolor[HTML]{EFEFEF}\ce{LiH}, 12 qubits (sto6g, JW)} \\ \hline
CS           & $2.32 \times 10^7$ & $531441$  & (50, 58, 43)  & $8.81 \times 10^3$ & $2910$ & $2.24 \times 10^5$  & $2.36 \times 10^5$ \\ \hline
LBCS         & $5.62 \times 10^5$ & $128002$ &  (2, 27, 1)    & $\mathbf{2.15 \times 10^2}$ & $204$ & $1.71 \times 10^4$  &  $1.75 \times 10^4$\\ \hline
Dec. Diag.   & $8.82 \times 10^4$ & $802$ & (32, 693, 2)      & $2.56 \times 10^4$ & $6.99$ & $3.22 \times 10^2$ & $2.59 \times 10^4$ \\ \hline
Derand. CS   & $\mathbf{7.13 \times 10^4}$ & $1814$ & (4, 433, 1)      & $7.13 \times 10^3$ & $4.89$ & $1.09 \times 10^2$  & $\mathbf{7.25 \times 10^3}$ \\ \hline
Derand. LBCS & $8.78 \times 10^4$ & $976$ & (11, 514, 1)      & $8.76 \times 10^3$ & $4.55$ & $7.86 \times 10^1$  & $8.84 \times 10^3$ \\ \hline
Derand. DD   & $9.32 \times 10^4$ & $\mathbf{243}$ & (462, 468, 1)     & $3.17 \times 10^4$ & $\mathbf{4.51}$ & $\mathbf{6.45 \times 10^1}$  & $3.18 \times 10^4$ \\ \hline
% Dec Diag: 2.56e4 (for opt DD) + 1.43e1 (sampling)
% Derand DD: 2.56e4 (for opt DD) + 6.08e3 (sampling)
\hline
\multicolumn{8}{|c|}{\cellcolor[HTML]{EFEFEF}\ce{N_2}, 16 qubits (sto6g, JW)} \\ \hline
% & LBCS         &  &  &   &  &  &   &  \\ \cline{2-9} 
Dec. Diag.   & $2.36 \times 10^6$ & $1134$  & (536, 11084, 9)  & $\mathbf{1.39 \times 10^5}$ & $\mathbf{123}$ & $1.81 \times 10^4$  & $\mathbf{1.57 \times 10^5}$ \\ \hline
Derand. CS   & $3.88 \times 10^6$ & $10015$ & (778, 35225, 1)  & $7.18 \times 10^5$ & $310$ & $7.84 \times 10^4$  & $7.97 \times 10^5$  \\ \hline
Derand. LBCS & $\mathbf{2.07 \times 10^6}$ & $5628$ & (73, 10274, 1)  & $3.91 \times 10^5$ & $173$ & $2.83 \times 10^4$  & $4.20 \times 10^5$ \\ \hline
Derand. DD   & $3.48 \times 10^6$ & $\mathbf{488}$ & (8714, 9754, 2)  & $5.31 \times 10^5$ & $140$ & $\mathbf{5.95 \times 10^3}$  & $5.37 \times 10^5$ \\ \hline
% Dec Diag: 1.39e5 (for opt DD) + 4.89e2 (sampling)
% Derand DD: 1.39e5 (for opt DD) + 3.92e5 (sampling)
\end{tabular}
}
\caption{
Comparison of resource utilization with experiments on a classical simulator by different measurement methods in estimating $\Tr(\rho H)$ to achieve an accuracy of $5\times 10^{-3}$ Hartree, with $H$ set to be different different molecular Hamiltonians (Table~\ref{table:molecules}) and $\rho$ as the ground state. Values shown are averages over $192$ independent runs against different metrics. Metrics shown are described in Section~\ref{sec:performance_metrics_benchmark}. Note that the contribution to classical post-processing time is from the estimator which is set to be the Bayesian estimator for all molecules except for \ce{N_2} where it is set to be the MC estimator. Lowest values obtained for each metric against a Hamiltonian is boldfaced.}
\label{tab:resource_analysis_sim}
\end{table}

\begin{table}[H]
\centering
\footnotesize
{\renewcommand{\arraystretch}{1.2}
% \begin{tabular}{|c|l|l|r|l|l|l|l|l|l|}
\begin{tabular}{|l||l|r|r|r|r|}
\hline
\multicolumn{1}{|c|}{\multirow{2}{*}{\begin{tabular}[c]{@{}c@{}}Measurement \\ Method\end{tabular}}} & 
\multicolumn{1}{c|}{\multirow{2}{*}{\begin{tabular}[c]{@{}c@{}} Predicted quantum\\ runtime [s]\end{tabular}}} & \multicolumn{4}{c|}{log(R)} \\ \cline{3-6} 
\multicolumn{1}{|c|}{} &
\multicolumn{1}{c|}{} &
\multicolumn{1}{c|}{A} &
\multicolumn{1}{c|}{B} &
\multicolumn{1}{c|}{C} &
\multicolumn{1}{c|}{D} \\ \hline
\hline 
\multicolumn{6}{|c|}{\cellcolor[HTML]{EFEFEF}\ce{H_2}, 5 qubits (3-21g, JW)} \\ \hline
CS           & $2.4 \times 10^2$ & $5.9$ & $10.5$ & $15.4$ & $20.2$ \\ \hline
LBCS         & $2.6 \times 10^2$ & $5.9$ & $10.6$ & $15.4$ & $20.3$ \\ \hline
Dec. Diag.   & $9.0 \times 10^1$ & $\mathbf{4.9}$ & $9.5$ & $14.4$ & $19.2$ \\ \hline
Derand. CS  & $7.3 \times 10^1$ & $7.5$ & $9.4$ & $14.2$ & $19.0$ \\ \hline
Derand. LBCS & $\mathbf{5.8 \times 10^1}$ & $7.8$ & $\mathbf{9.3}$ & $\mathbf{14.0}$ & $\mathbf{18.8}$ \\ \hline
Derand. DD   & $7.6 \times 10^1$ & $7.6$ & $9.5$ & $14.2$ & $19.1$ \\ \hline
\hline
\multicolumn{6}{|c|}{\cellcolor[HTML]{EFEFEF}\ce{HeH^+}, 6 qubits (6-31g, JW)} \\ \hline
CS           & $4.1 \times 10^2$ & $6.4$ & $11.0$ & $15.9$ & $20.8$ \\ \hline
LBCS         & $3.2 \times 10^2$ & $\mathbf{6.1}$ & $10.8$ & $15.7$ & $20.5$ \\ \hline
Dec. Diag.   & $\mathbf{1.1 \times 10^2}$ & $6.5$ & $\mathbf{9.7}$ & $\mathbf{14.6}$ & $\mathbf{19.4}$ \\ \hline
Derand. CS  & $1.2 \times 10^2$ & $8.6$ & $10.0$ & $14.7$ & $19.5$ \\ \hline
Derand. LBCS & $1.6 \times 10^2$ & $8.9$ & $10.3$ & $15.0$ & $19.8$ \\ \hline
Derand. DD   & $1.5 \times 10^2$ & $8.7$ & $10.3$ & $14.9$ & $19.8$ \\ \hline
\hline
\multicolumn{6}{|c|}{\cellcolor[HTML]{EFEFEF}\ce{HeH^+}, 8 qubits (6-31g, JW)} \\ \hline
CS           & $2.8 \times 10^3$ & $8.3$ & $12.9$ & $17.8$ & $22.7$ \\ \hline
LBCS         & $5.0 \times 10^2$ & $\mathbf{6.6}$ & $11.2$ & $16.1$ & $20.9$ \\ \hline
Dec. Diag.   & $1.5 \times 10^2$ & $8.2$ & $10.1$ & $14.9$ & $19.7$ \\ \hline
Derand. CS  & $1.3 \times 10^2$ & $9.1$ & $10.3$ & $14.8$ & $19.6$ \\ \hline
Derand. LBCS & $\mathbf{1.0 \times 10^2}$ & $8.9$ & $\mathbf{10.0}$ & $\mathbf{14.5}$ & $\mathbf{19.4}$ \\ \hline
Derand. DD   & $1.2 \times 10^2$ & $9.2$ & $10.2$ & $14.7$ & $19.5$ \\ \hline
\hline
\multicolumn{6}{|c|}{\cellcolor[HTML]{EFEFEF}\ce{LiH}, 12 qubits (sto6g, JW)} \\ \hline
CS           & $1.2 \times 10^4$ & $10.1$ & $14.4$ & $19.3$ & $24.1$ \\ \hline
LBCS         & $2.8 \times 10^2$ & $\mathbf{6.6}$ & $10.7$ & $15.5$ & $20.4$ \\ \hline
Dec. Diag.   & $4.4 \times 10^1$ & $10.2$ & $10.4$ & $13.7$ & $18.5$ \\ \hline
Derand. CS  & $\mathbf{3.6 \times 10^1}$ & $8.9$ & $\mathbf{9.4}$ & $\mathbf{13.5}$ & $\mathbf{18.3}$ \\ \hline
Derand. LBCS & $4.4 \times 10^1$ & $9.1$ & $9.6$ & $13.7$ & $18.5$ \\ \hline
Derand. DD   & $4.7 \times 10^1$ & $10.4$ & $10.6$ & $13.8$ & $18.6$ \\ \hline
\hline
\multicolumn{6}{|c|}{\cellcolor[HTML]{EFEFEF}\ce{N_2}, 16 qubits (sto6g, JW)} \\ \hline
Dec. Diag.   & $1.2 \times 10^3$ & $\mathbf{11.9}$ & $\mathbf{12.7}$ & $17.0$ & $21.8$ \\ \hline
Derand. CS  & $1.9 \times 10^3$ & $13.5$ & $13.8$ & $17.5$ & $22.3$ \\ \hline
Derand. LBCS & $1.0 \times 10^3$ & $12.9$ & $13.2$ & $\mathbf{16.9}$ & $\mathbf{21.7}$ \\ \hline
Derand. DD   & $1.7 \times 10^3$ & $13.2$ & $13.6$ & $17.4$ & $22.2$ \\ \hline
\end{tabular}
}
\caption{Comparison of predicted resource utilization on IBM Quantum devices by different measurement methods in estimating $\Tr(\rho H)$ to achieve an accuracy of $5\times 10^{-3}$ Hartree, with $H$ set to be different molecular Hamiltonians (Table~\ref{table:molecules}) and $\rho$ is the ground state. Predicted quantum runtime or the wall-clock that would be spent by the quantum device in executing experiments is shown for the different measurement methods based on the number of shots required to achieve $5\times 10^{-3}$ Hartree (Table~\ref{tab:resource_analysis_sim}) and assuming the quantum runtime would primarily be from delay between circuit executions. Estimates of resource utilization, indicated by $R$, here account for classical and predicted quantum computing resources via the heuristic in Eq.~\ref{eq:resource_utilization_heuristic} and using the regimes of Table~\ref{table:weights_heuristic}. Lowest values obtained for each metric against a Hamiltonian is boldfaced.}
\label{tab:resource_analysis_sim_predicted_for_dev}
\end{table}

\subsection{Experiments on quantum device} \label{sec:device_results}
% Convergence studies of RMSE with number of shots and tables of other metrics for finite budget of queries. 
As in the case of our experiments on the classical simulator, we consider the molecular Hamiltonians listed in Table~\ref{table:molecules}. We set the quantum state $\rho$ to be a random ansatz as discussed in Section~\ref{subsec:expt_setup}. This avoids the significant overhead of conducting a large VQE experiment to simulate the ground state of the Hamiltonian, and instead is representative of a typical state that might appear midway through a VQE optimization.

We carry out a comparison of all the candidate measurement methods for the smaller tapered Hamiltonians~(\ref{table:molecules}), but consider a subset of the measurement methods for the larger sized molecules. As the number of qubits $n$ of the Hamiltonian $H$ increases, the number of unique measurement bases queried by uniform CS or LBCS can become as high as our measurement budget which become too expensive to run in experiments on IBM quantum devices given current classical latencies (Section~\ref{sec:performance_metrics_benchmark}) and experimental constraints (Section~\ref{subsec:expt_setup}). We thus limit the number of circuits by considering the methods of decision diagrams and derandomization on the larger sized molecules.

\begin{figure}[H]
    \centering
    \includegraphics[width=0.6\textwidth]{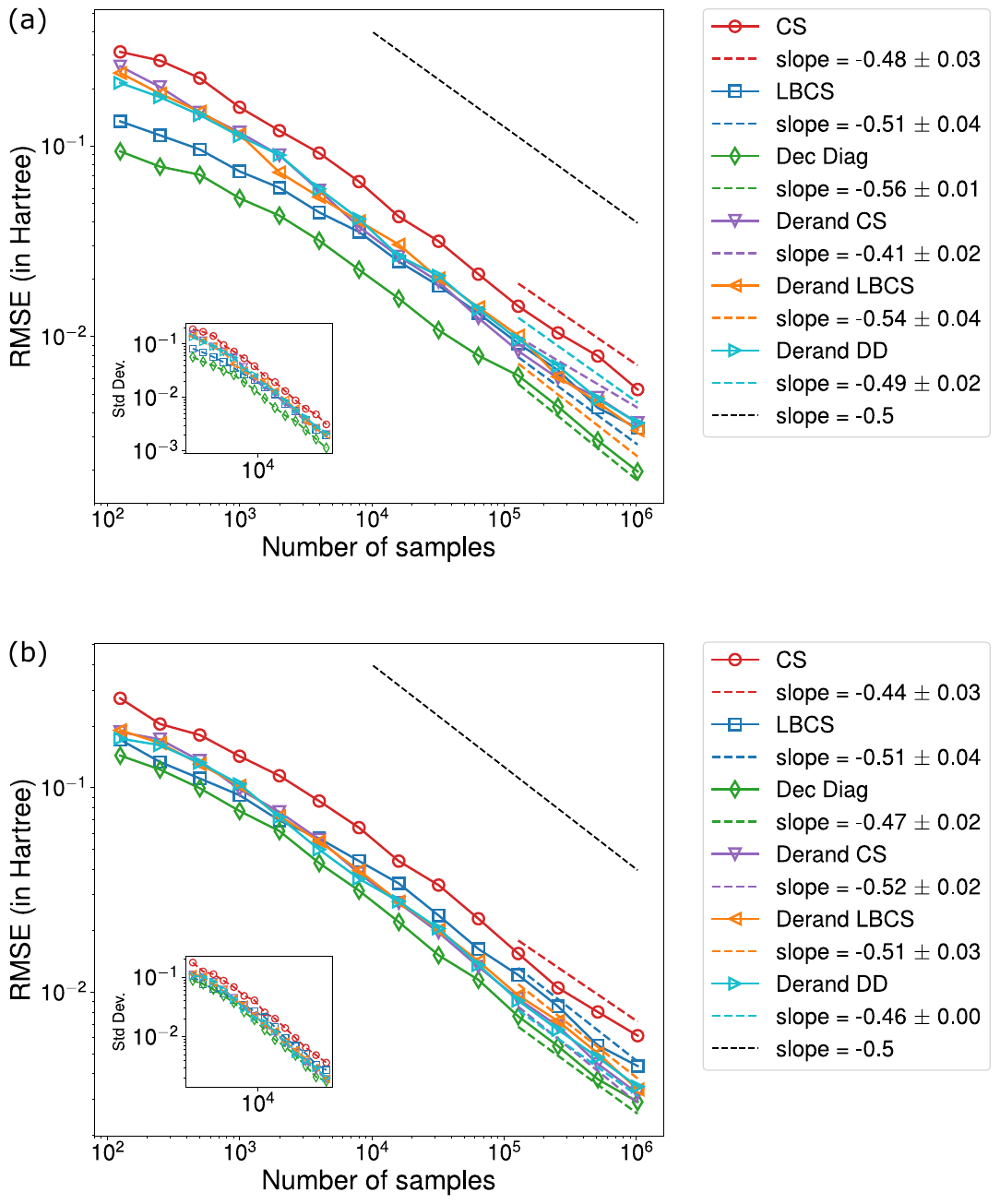}
    \caption{Comparison of empirical RMSE achieved in experiments on quantum device by different measurement methods in estimating $\Tr(\rho H)$ with $\rho$ set as a randomly fixed ansatz and $H$ is the Hamiltonian of (a) tapered \ce{H_2} ($5$ qubits, 3-21g basis, JW encoding), and (b) tapered \ce{HeH^{+}} ($6$ qubits, 3-21g basis, JW encoding). RMSE is shown with the number of samples (or shots) made. The estimator for each measurement method is set to be the Bayesian estimator.}
    \label{fig:comparison_meas_methods_H2_5q_HeH_6q}
\end{figure}

\paragraph{Tapered Hamiltonians.} We show convergence results for the tapered Hamiltonian \ce{H_2} (3-21g, JW, 5 qubits) in Figure~\ref{fig:comparison_meas_methods_H2_5q_HeH_6q}(a) and tapered Hamiltonian \ce{HeH^{+}} (3-21g, JW, 6 qubits) in Figure~\ref{fig:comparison_meas_methods_H2_5q_HeH_6q}(b). In these experimental results, we find larger differences in performance of the measurement methods than we observed earlier in the simulated results in Figure~\ref{fig:comparison_meas_methods_H2_5q_HeH_6q_sim}. The notable change from our observations on the simulator is that derandomization methods perform poorly compared to decision diagrams. This could be due to effects of measurement noise and choosing hyperparameters for derandomization via tuning on the simulator instead of the quantum device. Overall in the case of tapered \ce{H_2}, we can reduce the number of samples required to achieve an accuracy of $5$ milli Hartree by $57\%$ by switching from Derand. CS to DD, by $57\%$ as well by switching from LBCS to DD, and by around $85\%$ by opting for DD instead of CS. In the case of tapered \ce{HeH^{+}}, the resource reduction in the number of samples achieved by DD is more than $83\%$ compared to CS, $57\%$ compared to LBCS, and more than $30\%$ compared to any derandomization method.

\begin{figure}[H]
    \centering
    \includegraphics[width=0.6\textwidth]{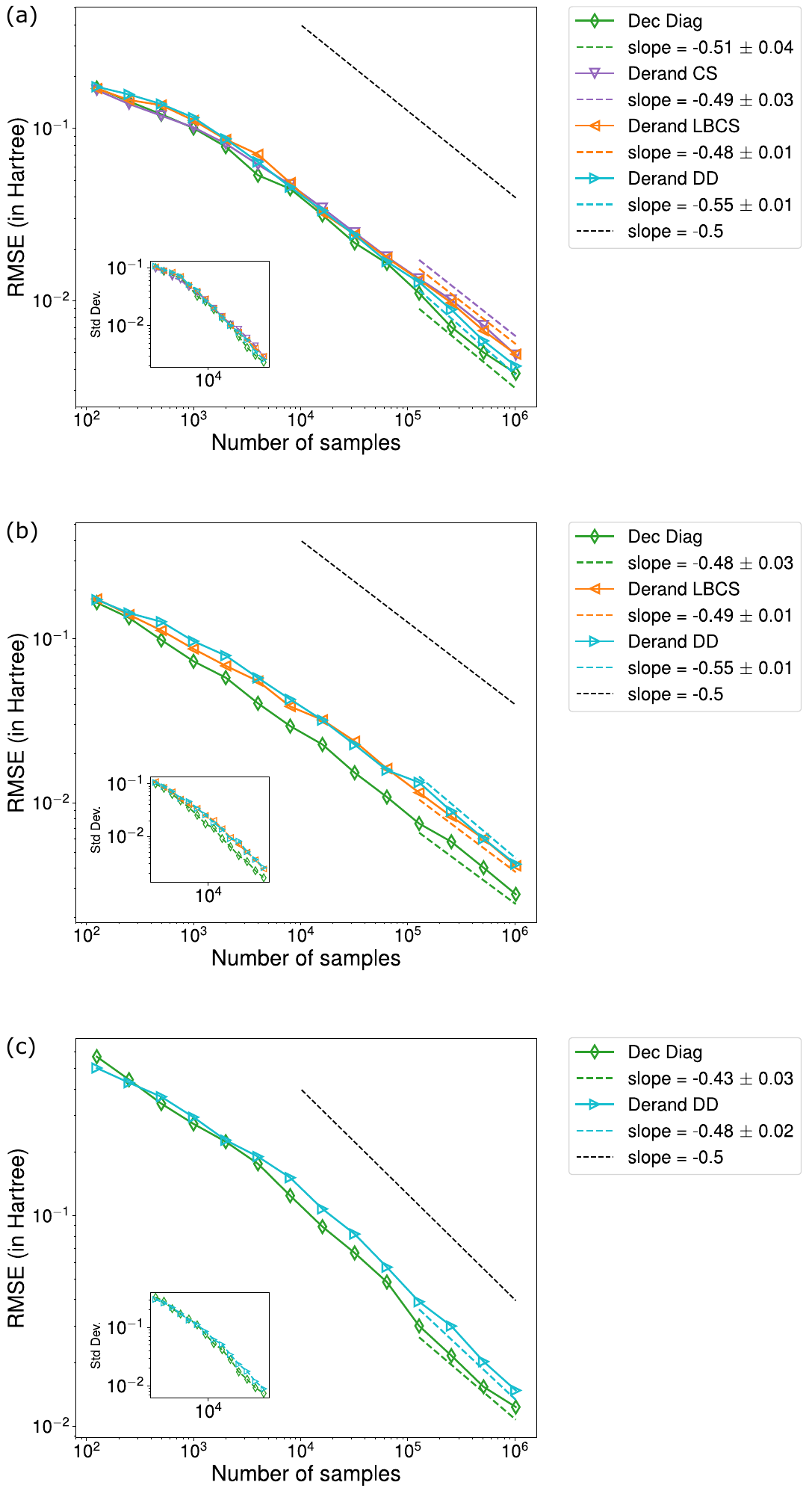}
    \caption{Comparison of empirical RMSE achieved in experiments on a quantum device by different measurement methods in estimating $\Tr(\rho H)$ with $\rho$ set as a randomly fixed ansatz and $H$ is the Hamiltonian of (a) \ce{HeH^{+}} ($8$ qubits, 6-31g basis, JW encoding), (b) \ce{LiH} ($12$ qubits, sto-6g basis, JW encoding), and (c) \ce{N_2} ($16$ qubits, sto-6g basis, JW encoding). RMSE is shown with the number of samples (or shots) made. The estimator for each measurement method is set to be the Bayesian estimator.}
    \label{fig:comparison_meas_methods_HeH_8q_LiH_12q_N2_16q}
\end{figure}

\paragraph{8 to 16 qubit Hamiltonians.} We show results for \ce{HeH^{+}} (6-31g, JW, $8$ qubits) in Figure~\ref{fig:comparison_meas_methods_HeH_8q_LiH_12q_N2_16q}(a), \ce{LiH} (sto6g, JW, $12$ qubits) in Figure~\ref{fig:comparison_meas_methods_HeH_8q_LiH_12q_N2_16q}(b), and \ce{N_2} in Figure~\ref{fig:comparison_meas_methods_HeH_8q_LiH_12q_N2_16q}(c). Note that as mentioned earlier, we benchmark the more frugal methods against these molecular Hamiltonians to keep the number of quantum compilations required during experiments reasonable and latencies incurred from circuit loading low. Across the three molecular Hamiltonians, we again find that optimized decision diagrams (Dec. Diag.) outperform derandomization (Derand.) methods. In particular for \ce{LiH}, decision diagrams are able to achieve an accuracy of $5$ milli-Hartree with around $40\%$ fewer shots than Derand. LBCS. On \ce{N_2}, the difference between DD and derandomization reduces from before with DD achieving a resource reduction of only $25\%$ at $5$ milli Hartree. Overall, DD consistently requires fewer measurements compared to derandomization methods.

\paragraph{Resource utilization.}
In Table~\ref{tab:resource_analysis_dev}, we report the resource utilization for estimating $\Tr(\rho H)$ within an accuracy cutoff of $5$ milli Hartree. We notice that across the board, Derand. DD requires the least number of unique circuits to be executed, as expected from our results on the simulator. Taking into account both quantum and classical computational resource utilization, decision diagrams are preferred over derandomization. It requires fewer shots to reach the same accuracy in experiments on the device and less classical computational runtime in generating samples. This is also supported by the weighted resource utilization reported in Table~\ref{tab:resource_analysis_dev} based on the heuristic and different regimes introduced in Section~\ref{sec:performance_metrics_benchmark}.

\begin{table}[ht!]
\centering
\footnotesize
{\renewcommand{\arraystretch}{1.2}
% \begin{tabular}{|c|l|l|r|l|l|l|l|l|l|}
\begin{tabular}{|l||l|r|r|l|r|l|l|r|r|r|r|}
\hline
% \begin{tabular}[c]{@{}c@{}}Molecular\\ Hamiltonian\end{tabular}& 
\multicolumn{1}{|c|}{\multirow{2}{*}{\begin{tabular}[c]{@{}c@{}}Measurement \\ Method\end{tabular}}} & 
\multicolumn{1}{c|}{\multirow{2}{*}{\begin{tabular}[c]{@{}c@{}}\# shots\\ required\end{tabular}}}    & 
\multicolumn{1}{c|}{\multirow{2}{*}{\begin{tabular}[c]{@{}c@{}}\# unique\\ circuits \end{tabular}}}  & 
\multicolumn{1}{c|}{\multirow{2}{*}{\begin{tabular}[c]{@{}c@{}}Median \# shots per circ.\\ {\scriptsize(all, top 5\%, bottom 5\%)}\end{tabular}}} & 
\multicolumn{3}{c|}{Classical runtime [s]} &
% \multicolumn{1}{c|}{\multirow{2}{*}{\begin{tabular}[c]{@{}c@{}}Classical \\ \\ runtime [s]\end{tabular}}} & 
% \multicolumn{1}{c|}{\multirow{2}{*}{\begin{tabular}[c]{@{}c@{}}Classical \\ post-process.\\ runtime [s]\end{tabular}}} &
% \multicolumn{1}{c|}{\multirow{2}{*}{\begin{tabular}[c]{@{}c@{}}Classical\\ latencies [s]\end{tabular}}} &
% \multicolumn{1}{c|}{\begin{tabular}[c]{@{}c@{}}Classical\\ quantum\\ overhead\end{tabular}} & 
\multicolumn{1}{c|}{\multirow{2}{*}{\begin{tabular}[c]{@{}c@{}}Quantum\\ runtime [s]\end{tabular}}} & \multicolumn{4}{c|}{log(R)} \\ \cline{5-7} \cline{9-12} 
\multicolumn{1}{|c|}{} &
\multicolumn{1}{c|}{} &
\multicolumn{1}{c|}{} &
\multicolumn{1}{c|}{} &
\multicolumn{1}{c|}{pre-process.} &
\multicolumn{1}{c|}{post-process.} &
\multicolumn{1}{c|}{latencies} &
\multicolumn{1}{c|}{} &
\multicolumn{1}{c|}{A} &
\multicolumn{1}{c|}{B} &
\multicolumn{1}{c|}{C} &
\multicolumn{1}{c|}{D} \\ \hline
\hline 
\multicolumn{12}{|c|}{\cellcolor[HTML]{EFEFEF}\ce{H_2}, 5 qubits (3-21g, JW)} \\ \hline
CS           & $1.2 \times 10^6$ & $243$  & (4917, 5056, 4788) & $2.3 \times 10^2$ & $6.5$  & $8.7 \times 10^4$  & $6.1 \times 10^2$ & $6.7$ & $11.4$ & $16.3$ & $21.1$ \\ \hline
LBCS         & $4.2 \times 10^5$ & $243$  &  (769, 10081, 76)  & $8.0 \times 10^1$ & $3.3$ & $3.0 \times 10^4$  & $2.1 \times 10^2$ & $5.7$ & $10.4$ & $15.3$ & $20.1$ \\ \hline
Dec. Diag.   & $\mathbf{1.8 \times 10^5}$ & $43$ & (1371, 15526, 37)    & $\mathbf{4.3 \times 10^1}$ & $\mathbf{1.1}$ & $\mathbf{1.3 \times 10^4}$  & $\mathbf{9.3 \times 10^1}$ & $\mathbf{4.9}$ & $\mathbf{9.5}$ & $\mathbf{14.4}$ & $\mathbf{19.3}$ \\ \hline
Derand. CS   & $4.3 \times 10^5$ & $49$ & (15740, 16442, 1)    & $9.7 \times 10^3$ & $10.1$ & $3.1 \times 10^4$  & $2.2 \times 10^2$ & $9.2$ & $10.6$ & $15.3$ & $20.1$ \\ \hline
Derand. LBCS & $4.1 \times 10^5$ & $55$ & (28, 15887, 1)       & $9.9 \times 10^3$ & $9.3$ & $3.0 \times 10^4$  & $2.1 \times 10^2$ & $9.2$ & $10.6$ & $15.3$ & $20.1$ \\ \hline
Derand. DD   & $4.9 \times 10^5$ & $\mathbf{32}$ & (17429, 17441, 1) & $8.5 \times 10^3$ & $8.7$ & $3.5 \times 10^4$  & $2.5 \times 10^2$ & $9.1$ & $10.7$ & $15.4$ & $20.2$ \\ \hline 
% Dec Diag: 2.88e1 (for opt DD) + 1.38e1 (sampling)
% Derand DD: 2.88e1 (for opt DD) + 8.46e3 (sampling)
\hline
\multicolumn{12}{|c|}{\cellcolor[HTML]{EFEFEF}\ce{HeH^+}, 6 qubits (6-31g, JW)} \\ \hline
CS           & $1.8 \times 10^6$ & $729$ & (2525, 2731, 2347)  & $3.5 \times 10^2$  & $6.4$ & $1.4 \times 10^5$  & $9.3 \times 10^2$ 
& $7.2$ & $11.8$ & $16.7$ & $21.6$ \\ \hline
LBCS         & $7.1 \times 10^5$ & $728$ & (418, 4566, 42)  & $\mathbf{1.4 \times 10^2}$ & $3.9$ & $5.3 \times 10^4$  & $3.6 \times 10^2$ 
& $\mathbf{6.2}$ & $10.9$ & $15.8$ & $20.6$ \\ \hline
Dec. Diag.   & $\mathbf{3.0 \times 10^5}$ & $234$ &  (355, 8071, 10) & $5.7 \times 10^2$ & $\mathbf{1.4}$ & $\mathbf{2.2 \times 10^4}$  & $\mathbf{1.5 \times 10^2}$ 
& $6.6$ & $\mathbf{10.0}$ & $\mathbf{14.9}$ & $\mathbf{19.7}$ \\ \hline
Derand. CS   & $4.3 \times 10^5$ & $189$ & (1009, 5071, 1) & $9.8 \times 10^3$ & $10.1$ & $3.2 \times 10^4$  & $2.2 \times 10^2$ 
& $9.2$ & $10.7$ & $15.3$ & $20.1$ \\ \hline
Derand. LBCS & $5.0 \times 10^5$ & $185$ &  (1796, 5824, 1) & $1.2 \times 10^4$ & $9.8$ & $3.8 \times 10^4$  & $2.6 \times 10^2$ 
& $9.4$ & $10.8$ & $15.5$ & $20.3$ \\ \hline
Derand. DD   & $4.4 \times 10^5$ & $\mathbf{124}$ & (5157, 5179, 1) & $8.5 \times 10^3$ & $8.7$ & $3.3 \times 10^4$  & $2.2 \times 10^2$ 
& $9.1$ & $10.6$ & $15.3$ & $20.1$ \\ \hline
% Dec Diag: 5.45e2 (for opt DD) + 2.50e1 (sampling)
% Derand DD: 5.45e2 (for opt DD) + 7.95e3 (sampling)
\hline
\multicolumn{12}{|c|}{\cellcolor[HTML]{EFEFEF}\ce{HeH^+}, 8 qubits (6-31g, JW)} \\ \hline
Dec. Diag.   & $\mathbf{5.5 \times 10^5}$ & $787$ & (181, 5392, 4) & $\mathbf{3.6 \times 10^3}$  & $\mathbf{5.3}$ & $\mathbf{4.1 \times 10^4}$  & $\mathbf{2.8 \times 10^2}$ 
& $\mathbf{8.3}$ & $\mathbf{10.7}$ & $\mathbf{15.5}$ & $\mathbf{20.4}$ \\ \hline
Derand. CS   & $1.0 \times 10^6$ & $548$ & (59, 9254, 2)  & $3.5 \times 10^4$ & $41.5$ & $7.5 \times 10^4$  & $5.1 \times 10^2$ 
& $10.5$ & $11.6$ & $16.1$ & $21.0$ \\ \hline
Derand. LBCS & $1.0 \times 10^6$ & $518$ & (25, 9125, 2) & $3.5 \times 10^4$ & $3 4.1$ & $7.5 \times 10^4$  & $5.1 \times 10^2$ 
& $10.5$ & $11.6$ & $16.1$ & $21.0$ \\ \hline
Derand. DD   & $7.2 \times 10^5$ & $\mathbf{210}$ & (3971, 6042, 1) & $2.3 \times 10^4$ & $28.7$ & $5.4 \times 10^4$  & $3.7 \times 10^2$ 
& $10.1$ & $11.3$ & $15.8$ & $20.6$ \\ \hline
% Dec Diag: 3.54e3 (for opt DD) + 5.84e1 (sampling)
% Derand DD: 3.54e3 (for opt DD) + 1.94e4 (sampling)
\hline
\multicolumn{12}{|c|}{\cellcolor[HTML]{EFEFEF}\ce{LiH}, 12 qubits (sto6g, JW)} \\ \hline
Dec. Diag.   & $\mathbf{3.2 \times 10^5}$  & $811$ & (112, 2504, 7)  & $\mathbf{2.6 \times 10^4}$ & $\mathbf{12.8}$ & $\mathbf{2.4 \times 10^4}$  & $\mathbf{1.6 \times 10^2}$ 
& $\mathbf{10.2}$ & $\mathbf{10.8}$ & $\mathbf{15.0}$ & $\mathbf{19.8}$ \\ \hline
Derand. LBCS & $7.0 \times 10^5$ & $979$ & (23, 4090, 2) & $6.9 \times 10^4$ & $13.0$ & $5.2 \times 10^4$  & $3.5 \times 10^2$ 
& $11.1$ & $11.7$ & $15.8$ & $20.6$ \\ \hline
Derand. DD   & $7.5 \times 10^5$ & $\mathbf{284}$ & (3715, 3727, 1) & $7.4 \times 10^4$  & $13.0$  & $5.6 \times 10^4$  & $3.8 \times 10^2$ 
& $11.2$ & $11.8$ & $15.9$ & $20.7$ \\ \hline
% Dec Diag: 2.56e4 (for opt DD) + 5.20e1 (sampling)
% Derand DD: 2.56e4 (for opt DD) + 4.85e4 (sampling)
\hline
\multicolumn{12}{|c|}{\cellcolor[HTML]{EFEFEF}\ce{N_2}, 16 qubits (sto6g, JW)} \\ \hline
Dec. Diag.   & $\mathbf{8.9 \times 10^6}$ & $1137$ & (15298, 40070, 480) & $\mathbf{1.4 \times 10^5}$ & $516$ & $\mathbf{6.5 \times 10^5}$  & $\mathbf{4.5 \times 10^3}$ 
& $\mathbf{11.9}$ & $\mathbf{13.6}$ & $\mathbf{18.3}$ & $\mathbf{23.1}$ \\ \hline
Derand. DD   & $1.2 \times 10^7$ & $\mathbf{488}$ & (33477, 94988, 2) & $1.6 \times 10^6$ & $369$ & $8.8 \times 10^5$ & $6.2 \times 10^3$ 
& $14.3$ & $14.7$ & $18.6$ & $23.5$ \\ \hline
% Dec Diag: 1.39e5 (for opt DD) + 1.99e3 (sampling)
% Derand DD: 1.39e5 (for opt DD) + 1.43e6 (sampling)
\end{tabular}
}
\caption{Comparison of resource utilization with experiments on IBM Quantum devices by different measurement methods in estimating $\Tr(\rho H)$ to achieve an accuracy of $5\times 10^{-3}$ Hartree, with $H$ set to be different molecular Hamiltonians (Table~\ref{table:molecules}) and $\rho$ is a random ansatz.  Values shown are averages over $192$ bootstrapped runs against different metrics. Metrics shown are described in Section~\ref{sec:performance_metrics_benchmark}. Number of unique circuits run contribute to classical latencies such as compilation. Number of shots per circuit (all, top 5\%, bottom 5\%) summarizes shot distribution across circuits and is an indication of latencies due to circuit loading on control electronics. Classical pre-processing runtime is the wall-clock time spent by measurement methods in constructing a query distribution and generating samples. Classical post-processing runtime is the wall-clock time spent by the Bayesian estimator in computing an estimate from acquired measurement outcomes. Quantum runtime is the wall-clock spent by the quantum device in executing experiments. Estimates of resource utilization, indicated by $R$, account for both classical and quantum computing resources via the heuristic in Eq.~\ref{eq:resource_utilization_heuristic} and using the regimes of Table~\ref{table:weights_heuristic}. Lowest values obtained for each metric against a Hamiltonian is boldfaced.}
\label{tab:resource_analysis_dev}
\end{table}

\section{Conclusion} \label{sec:conclusion}
In this paper, we propose \benchname~(\benchabbr)~for benchmarking measurement methods designed to solve the problem of estimating the expectation value $\Tr(\rho H)$ of a quantum Hamiltonian $H$ with respect to a quantum state $\rho$. We presented the various evaluation criteria and explained the importance of considering utilization of both quantum and classical resources for these hybrid measurement methods in addition to the performance metric of accuracy achieved in estimating $\Tr(\rho H)$. In practice, CSHOREBench is empirically efficient to run on states of random quantum ansatz with fixed depth. This makes it an attractive tool to select measurement methods for a given Hamiltonian before experiments are executed on a quantum device. In numerical simulations on molecular Hamiltonians, we found that compact decision diagrams (DD) along with derandomization are the most competitive methods. In our experiments on IBM quantum devices, we observed that decision diagrams achieved a given accuracy with fewer quantum measurements than derandomization. However, derandomization of decision diagrams may be preferred if a low number of unique circuits is the most important consideration.

% Other domains besides VQE that have the measurement problem
The improvement in query complexity by using compact decision diagrams or derandomized decision diagrams for estimating $\Tr(\rho H)$ has multiple practical implications besides accelerating steps of variational quantum eigensolvers (VQE). For example, in the hybrid quantum-classical algorithm of Krylov subspace methods~\cite{parrish2019filterdiagonalization,motta2020determining}, the measurement problem appears when estimating matrix elements for the generalized eigenvalue problem and DDs may be advantageous in that setting as well, albeit optimized differently. 

% Adaptivity in queries (or distribution) as measurements are made
% Considering the case of low-depth or local-entangling measurements
There is still room for improvement in the design of measurement methods themselves. Randomized measurement methods could be improved by incorporating adaptive strategies during learning by building information about the quantum state $\rho$ under consideration in real-time in addition or instead of prior information. One tool to introduce adaptivity would be active learning which has been shown to be useful in practice for the learning tasks of quantum state tomography~\cite{nunn2010optimal} and Hamiltonian learning~\cite{dutt2023active}. The query complexity of randomized measurement methods could also be improved by allowing low-depth and locally-entangling measurements~\cite{ippoliti2023classical}. It is known that the query complexity is significantly improved~\cite{huang2020predicting} when global Clifford measurements are allowed but these require deeper circuits than one would typically want to expend on the measurement part of a near-term quantum algorithm.

% Future work on benchmarking
% Heuristics summarizing different benchmarking characteristics/criteria to rank different measurement methods
Aside from improving the measurement methods themselves, there is still potential for improving the benchmarking introduced here. We have proposed a heuristic comprising different regimes to summarize utilization of quantum and classical resources. Recommendations of heuristics for other platforms would aid in providing further guidance on which measurement methods to prefer for these platforms. Moreover, the list of measurement methods being benchmarked here could be extended to include grouping methods and other approaches~\cite{gresch2023guaranteed}. As for the instances considered as examples of common computations, future benchmarks would benefit from extending beyond minimal bases. We used minimal bases in this work in order to reduce the computational cost on the quantum side as far as possible, but these are not physically representative, so using correlation consistent bases would be preferable as we advance toward practically relevant quantum algorithms for chemistry.

\begin{acknowledgments}
The authors thank Andrew Eddins and Mirko Amico for helpful conversations regarding experiments on IBM quantum hardware. AD and ILC were supported in part by the U.S. Department of Energy, Office of Science, National Quantum Information Science Research Centers, and Co-Design Center for Quantum Advantage under contract DE-SC0012704. AD acknowledges the MIT SuperCloud and Lincoln Laboratory Supercomputing Center for providing computational resources.
\end{acknowledgments}

% \clearpage
%%%%%%%%%%%%%%%%%%%%%%%%%%%%%%%%%%%%%%%%%%%%%%%%%%%%%%%%%%%%%%%
%\bibliographystyle{alpha}
%\bibliographystyle{siam}

\bibliographystyle{IEEEtran}
\bibliography{randmeas_refs} % Produces the bibliography via BibTeX.

\clearpage
\appendix
{\begin{center}\textbf{\Large Supplementary Material: Practical Benchmarking of Randomized Measurements of Quantum Hamiltonians}\end{center}
\vspace{0.25in}}

In Section~\ref{app_sec:derand_DD}, we give details of the measurement method derandomization of decision diagrams which was introduced in Section~\ref{sec:derand} of the main paper. Specifically, we show how the associated cost function is obtained and how it may be computed efficiently on decision diagrams corresponding to molecular Hamiltonians. In Section~\ref{app_sec:details_molecules_and_meas}, we give details of the molecular Hamiltonians from Table~\ref{table:molecules} including their Pauli weight distributions and the query distributions obtained for these molecular Hamiltonians via LBCS or decision diagrams. Finally, we also comment on the convergence behavior of Adaptive Pauli Shadows (APS) as observed on the simulator.

\section{Details of derandomizing decision diagrams}\label{app_sec:derand_DD}
In this section of the Appendix, we give the technical details of derandomizing decision diagrams (Derand DD) which was proposed in Section~\ref{sec:derand} of the main paper. We will first show how the cost function for Derand DD is obtained by considering the general query distribution of the decision diagram to be $\beta$ and then show how the cost function can be computed efficiently on a decision diagram. As discussed in Section~\ref{sec:derand}, our starting point is the \textit{confidence bound} introduced in \cite{huang2021efficient}. For completeness, we show the proof in \cite[Lemma 2]{huang2021efficient} to motivate the confidence bound and then proceed to obtain the cost function for Derand DD.

\paragraph{Confidence bound on estimates.} Recall that we denote the estimates $\Tr(\rho Q^{(j)})$ as $\hat{\omega}^{(j)}$ and their true value as $\omega^{(j)}$. We analyze the probability of a large deviation of the estimates obtained from post-processing on the measurement outcomes against bases $\mathbf{B}$
\begin{align}
    \Pr\left[ \max_{j \in [L]} |\hat{\omega}^{(j)} - \omega^{(j)}| \geq \epsilon \right]
    &= \Pr\left[ \bigcup_{j \in [L]} |\hat{\omega}^{(j)} - \omega^{(j)}| \geq \epsilon \right] \\
    &\leq \sum_{j \in [L]}  \Pr\left[ |\hat{\omega}^{(j)} - \omega^{(j)}| \geq \epsilon \right] \quad \text{(Union bound)} \\
    &\leq 2 \sum_{j \in [L]}  \exp \left(-\frac{\epsilon^2}{2}  h(Q^{(j)};\mathbf{B}) \right) \quad \text{(Hoeffding's inequality)} \\
    &= 2 \sum_{j \in [L]}  \prod \limits_{s \in [M]} \exp \left(-\frac{\epsilon^2}{2}  \mathds{1}\{Q^{(j)} \triangleright B^{(s)}\} \right) \\
    &= 2 \sum_{j \in [L]}  \prod \limits_{s \in [M]} \left(1 - \eta \mathds{1}\{Q^{(j)} \triangleright B^{(s)}\} \right),
\end{align}
where $\eta = 1 - \exp(-\epsilon^2/2)$. We see that the probability of the estimate $\hat{\omega}^{(j)}$ deviating from the truth reduces exponentially with the number of measurement bases that hit/cover the target observable $Q^{(j)}$. We will call the upper bound without the constant factor (as derived above) of this probability as the confidence bound:
\begin{equation}
    \mathrm{CONF}_\epsilon(\mathbf{Q};\mathbf{B}) = \sum_{j \in [L]}  \prod \limits_{s \in [M]} \left(1 - \eta \mathds{1}\{Q^{(j)} \triangleright B^{(s)}\} \right).
\end{equation}

\paragraph{Expectation of the confidence bound.} To obtain the cost function for derandomization, it is desirable to compute the expectation of the confidence bound which is given by
\begin{align}
    \mathbb{E}\left[ \mathrm{CONF}_\epsilon(\mathbf{Q};\mathbf{B}) \right] &= \sum_{j \in [L]}  \prod \limits_{s \in [M]} \mathbb{E}\left[ \left(1 - \eta  \mathds{1}\{Q^{(j)} \triangleright B^{(s)}\}  \right)\right] \\
    &= \sum_{j \in [L]} \left(1 - \eta  \mathbb{E} \left[\mathds{1}\{Q^{(j)} \triangleright B^{(s)}\} \right]  \right)^M \\
    &= \sum_{j \in [L]} \left(1 - \eta \xi(Q^{(j)},\beta) \right)^M,
\end{align}
where the expectation is with respect to the query distribution $\beta$, we have used the fact that each $B^{(s)}$ is sampled independently and identically from the distribution $\beta$ for all $s \in [M]$, and where $\xi(Q^{(j)},\beta) = \Pr[Q^{(j)} \text{ covered by } \beta]$ denoting the coverage probability (notation introduced in Section~\ref{subsec:wmc_estimator}). If the query distribution $\beta=\prod_{k \in [n]} \beta_k $ is a product distribution (with the marginal distribution the $k$th qubit dented as $\beta_k$) as in the case of CS and LBCS, we have the following simplified expression
\begin{equation}
    \mathbb{E}\left[ \mathrm{CONF}_\epsilon(\mathbf{Q};\mathbf{B}) \right] = \sum_{j \in [L]} \left(1 - \eta \prod \limits_{k=1}^n \beta_k(Q^{(j)}_k)^{\mathds{1}\{Q^{(j)}_k \neq I\}}  \right)^M.
\end{equation}

\paragraph{Cost function for derandomization.} Let us now discuss how a cost function for derandomization of decision diagrams may be obtained using the confidence bound. Suppose $\mathbf{B}^{\#}$ contains the assignments of measurement bases for the first $(m-1)$ samples and first $k$ qubits of the $m$th measurement basis. We then have the following conditional expectation
\begin{align}
    \mathbb{E}\left[ \mathrm{CONF}_\epsilon(\mathbf{Q};\mathbf{B}) | \mathbf{B}^{\#} \right]
    &= \sum_{j \in [L]} \prod \limits_{m'=1}^{m-1} \left(1 - \eta \mathds{1}\{Q^{(j)} \triangleright {B^\#}^{(m')}\} \right)  \\
    &\times \left(1 - \eta \prod \limits_{k'=1}^{k} \mathds{1}\{Q^{(j)} \triangleright {B^\#}^{(m')}\} \Pr \left[Q^{(j)}_{k+1:n} \text{ covered by } \beta| {B^{\#}}^{(m)}_{1:k} \right] \right) \nonumber \\
    & \times \left(1 - \eta \Pr[Q^{(j)} \text{ covered by } \beta]  \right)^{M-m} \nonumber,
    \label{eq:conditional_expectation_conf_bound}
\end{align}
where we have denoted ${B^\#}^{(m)}$ as the $m$th measurement basis in $B^{\#}$, ${B^\#}^{(m)}_k$ as the $k$th qubit Pauli of the $m$th measurement basis, used the sub-scripted Pauli operator $Q^{(j)}_{k+1:n}$ to denote $Q^{(j)}_{k+1:n} = \otimes_{\ell=k+1}^n Q^{(j)}_\ell$ and similarly ${B^{\#}}^{(m)}_{1:k} = \otimes_{\ell = 1}^k {B^{\#}}^{(m)}_\ell$. In the above expression, we also have $\Pr \left[Q^{(j)}_{k+1:n} \text{ covered by } \beta| {B^{\#}}^{(m)}_{1:k} \right] = \xi(\otimes_{\ell=1}^k {B^{\#}}^{(m)}_\ell \otimes_{\ell'=k+1}^n Q^{(j)}_{\ell'}, \beta)$.

In the special case of the query distribution $\beta$ being a production distribution (e.g., as in LBCS), we have the following expression for the conditional expectation of the confidence bound
\begin{align}
    \mathbb{E}\left[ \mathrm{CONF}_\epsilon(\mathbf{Q};\mathbf{B}) | \mathbf{B}^{\#} \right] &= \sum_{j \in [L]} \prod \limits_{m'=1}^{m-1} \left(1 - \eta \mathds{1}\{Q^{(j)} \triangleright {B^\#}^{(m')}\} \right) \\
    &\times \left(1 - \eta \prod \limits_{k'=1}^{k} \mathds{1}\{Q^{(j)} \triangleright {B^\#}^{(m')}\} \prod \limits_{k'=k+1}^{n} \beta_{k'}(Q^{(j)}_{k'})^{\mathds{1}\{Q^{(j)}_{k'} \neq I\}} \right) \nonumber \\
    & \times \left(1 - \eta \prod \limits_{k=1}^n \beta_k(Q^{(j)}_k)^{\mathds{1}\{Q^{(j)}_k \neq I\}}  \right)^{M-m}. \nonumber
\end{align}

To choose the assignment of the $k$th qubit of the $m$th measurement basis, we then consider the following cost function
\begin{equation}
    {B_k^\#}^{(m)} = \argmin_{ W \in \{X,Y,Z\} } C(W) = \argmin_{ W \in \{X,Y,Z\} } \mathbb{E}\left[ \mathrm{CONF}_\epsilon(\mathbf{Q};\mathbf{B}) | \mathbf{B}^{\#}, B^{(m)}_k = W\right]
    \label{eq:derand_cost_function}
\end{equation}
where $\mathbf{B}^{\#}$ now corresponds to the assignments of measurement bases over the first $(m-1)$ samples and $(k-1)$ qubits of the $m$th measurement basis. Note that the above cost function requires the input of the experimental budget $M$. As was done in \cite{huang2021efficient} for derandomizing CS, we can remove the dependence on the measurement budget in the cost function by removing the third term in the product of Eq.~\ref{eq:conditional_expectation_conf_bound}.

Derandomization of a general query distribution $\beta$ is then given by Algorithm~\ref{algo:derandomization_DD} shown in the main part of the paper. 
%Let us now describe how the cost function of Eq.~\ref{eq:derand_cost_function} can be computed efficiently on a decision diagram for a molecular Hamiltonian.

\paragraph{Fast computation of the cost function on decision diagrams.}
As part of computing the cost function~\ref{eq:derand_cost_function} quickly, we need to be able to perform quick computation of the conditional probability $\Pr \left[Q^{(j)}_{k+1:n} \text{ covered by } \beta| {B^{\#}}^{(m)}_{1:k} \right]$ in Eq.~\ref{eq:conditional_expectation_conf_bound}. We will refer to this as the conditional probability of coverage.

To compute the conditional probability of coverage, let us now introduce some relevant notation and useful routines. We will refer to the decision diagram using its graph $G=(V,E)$ where we have denoted the set of nodes as $V$ and edges as $E$. We will refer to the directed edge from node $u$ to node $v$ as $E(u,v)$. As we mentioned in Section~\ref{sec:meas_methods}, each edge has a label associated with it, namely the Pauli decision being taken. This will be denoted by $P(u,v)$. 

For each node $v \in V$, we define the coverable set $C_v$ as the non-identity Pauli terms in $H$ that can be covered by measurements starting from the root of the decision diagram (which we denote as $r$ and for example, corresponds to node $0$ in the decision diagram of Figure~\ref{fig:viz_decision_diagrams_H2}(c)). For example, in Figure~\ref{fig:viz_decision_diagrams_H2}(c), the target observable $ZIII$ will be a part of the coverable set $C_9$ as the first qubit Pauli $Z$ is covered by the edge $E(0,9)$. We compute the coverable sets for each node recursively.

The coverable sets at the root node $r$ (or node $0$ in Figure~\ref{fig:viz_decision_diagrams_H2}) and at the terminal node $t$ (or node $-1$ in Figure~\ref{fig:viz_decision_diagrams_H2}) are set to be all the non-identity Pauli operators in the decomposition of the Hamiltonian $H$ i.e., $C_r = C_t := \mathbf{Q} \setminus I^{\otimes n}$. Suppose node $u$ is the child of node $r$, then $C_u = \{Q | Q \in C_r, \, Q_1 \in \{P(r,u),I\}\}$. We can do this for the nodes corresponding to the children of the root node but generally, nodes may have multiple parents. To obtain coverage for a general node $v$, we move down the decision diagram one layer of edges at a time and update the coverage of a node $v$ in layer $\ell$ based on its parents in the previous layer $\ell-1$ as $C_v = \bigcup_{u \in \mathrm{Parents}(v)} \{Q | Q \in C_u, \, Q_\ell \in \{P(u,v),I\}\}$.

Now, given that the current state of a measurement basis being proposed is at node $v$, we can compute the conditional probability $\Pr[Q_{k+1:n} \text{ covered by DD }| v]$ (where we have denoted the query distribution $\beta$ in Eq.~\ref{eq:conditional_expectation_conf_bound} by its DD) of covering any $Q \in C_v$ in a recursive fashion as
\begin{align}
    \Pr[Q_{k:n} \text{ covered by DD }| v, Q \in C_v] = \sum_{w \in \mathrm{Children}(w)} & \mathds{1}\{Q_k \triangleright P(v,w)\} \\ \nonumber
    & \times \Pr[P(v,w)] \Pr[Q_{k+1:n} \text{ covered by DD }| w, Q \in C_W],
\end{align}
where we have used $\Pr[P(v,w)]$ to denote the probability of taking the decision of Pauli measurement $P(v,w)$ that is available to us from the DD. We should note that there at most three children for any node $v$ corresponding to the three decisions of the single-qubit Paulis $\{X,Y,Z\}$. Finally, to compute the conditional probability of coverage $\Pr \left[Q^{(j)}_{k+1:n} \text{ covered by } \beta| {B^{\#}}^{(m)}_{1:k} \right]$, it is enough to note that making the sequence of measurements in ${B^{\#}}^{(m)}_{1:k}$ will place us at a node $v$ in the DD. This node $v$ will be unique. We can argue this by noting that if taking the decisions in ${B^{\#}}^{(m)}_{1:k}$ starting from the root node took us down to two different nodes, then these two nodes would have been merged from the initialization of the DD.

Overall, as part of the fast computation of the cost function in derandomization on DDs, we firstly move from the root node downwards to the terminal node to determine the node coverages $C_v$, and then secondly move from the terminal node upwards to the root node determine the conditional probabilities of coverage.

% \paragraph{Performance guarantee.}
% The derandomized LBCS algorithm is promised to do atleast as good as the randomized algorithm of LBCS. To see this, consider the first assignment:
% \begin{align}
%     {B_1^\#}^{(1)} &= \argmin_{ W \in \{X,Y,Z\} } \mathbb{E}\left[ \mathrm{CONF}_\epsilon(\mathbf{Q};\mathbf{B}) | B^{(1)}_1 = W\right] 
% \end{align}
% Note that we have the following
% \begin{align}
%     \min_{ W \in \{X,Y,Z\} } \mathbb{E}\left[ \mathrm{CONF}_\epsilon(\mathbf{Q};\mathbf{B}) | B^{(1)}_1 = W\right] &\leq \sum \limits_{ W \in \{X,Y,Z\} } \beta_1(W) \mathbb{E}\left[ \mathrm{CONF}_\epsilon(\mathbf{Q};\mathbf{B}) | B^{(1)}_1 = W\right] \\
%     &= \mathbb{E}\left[ \mathrm{CONF}_\epsilon(\mathbf{Q};\mathbf{B})\right]
% \end{align}
% % minimum < average (or expectation)

% Iteratively, we can then show that 
% \begin{equation}
%     \mathrm{CONF}_\epsilon(\mathbf{Q};\mathbf{B}^{\#}) \leq \mathbb{E}\left[ \mathrm{CONF}_\epsilon(\mathbf{Q};\mathbf{B})\right]
% \end{equation}
% where the former corresponds to the derandomized LBCS and the latter corresponds the randomized LBCS. This gives us a guarantee that the derandomized LBCS will perform at least as good as the randomized LBCS (in the average sense).

\section{Details of molecules and measurement methods}\label{app_sec:details_molecules_and_meas}
In this section of the Appendix, we describe the Pauli distributions of the different molecular Hamiltonians considered in Table~\ref{table:molecules} and illustrate query distributions obtained for these molecules via LBCS or DD. The Pauli decompositions of the Hamiltonians considered in this paper are available in our code repository.

\subsection{Pauli weight distributions}
The sample complexity of classical shadows~\cite{huang2020predicting} is known to be $O(3^w \log L)$ where $w$ is the maximum weight of any Pauli term in the Pauli decomposition of $H$. However, in practice, the dependence of the sample complexity on $w$ may be better for some measurement methods depending on mass of the query distribution on higher weight Paulis and the coefficients of higher weight Paulis in $H$.

\begin{figure}[ht!]
    \centering
    \includegraphics[width=\textwidth]{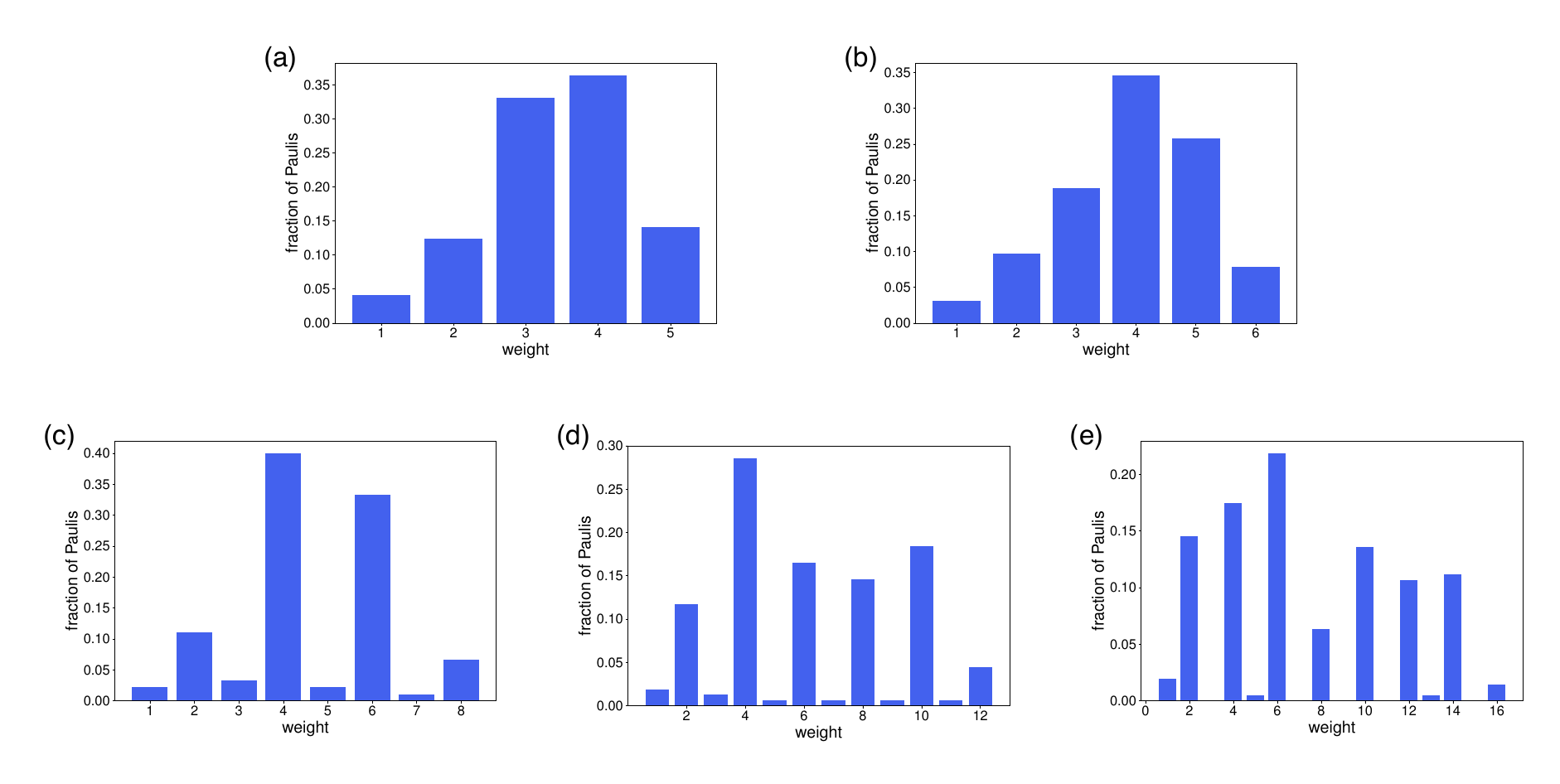}
    \caption{Pauli weight distributions across non-identity target Paulis in Hamiltonians of different molecules. Molecules correspond to those in Table~\ref{table:molecules} and as shown are (a) \ce{H_2}(5 qubits, 3-21g, JW), (b) \ce{HeH^{+}} (6 qubits, 3-21g, JW), (c) \ce{HeH^{+}} (8 qubits, 6-31g, JW), (d) \ce{LiH} (12 qubits, sto6g, JW), and (e) \ce{N_2} (16 qubits, sto6g, JW).}
    \label{fig:pauli_wt_distrn_molecules}
\end{figure}

Thus, it might be desirable to include Hamiltonians of the same size but with different Pauli weight distributions as part of benchmarking measurement methods. Here, in Figure~\ref{fig:pauli_wt_distrn_molecules}, we visualize the Pauli weight distributions of the Hamiltonians from Table~\ref{table:molecules} which range from unimodal to skewed to bimodal.

\subsection{Query distributions for different molecules}\label{app_sec:molecules_query_distrn}
\begin{figure}[ht!]
    \centering
    \includegraphics[width=\textwidth]{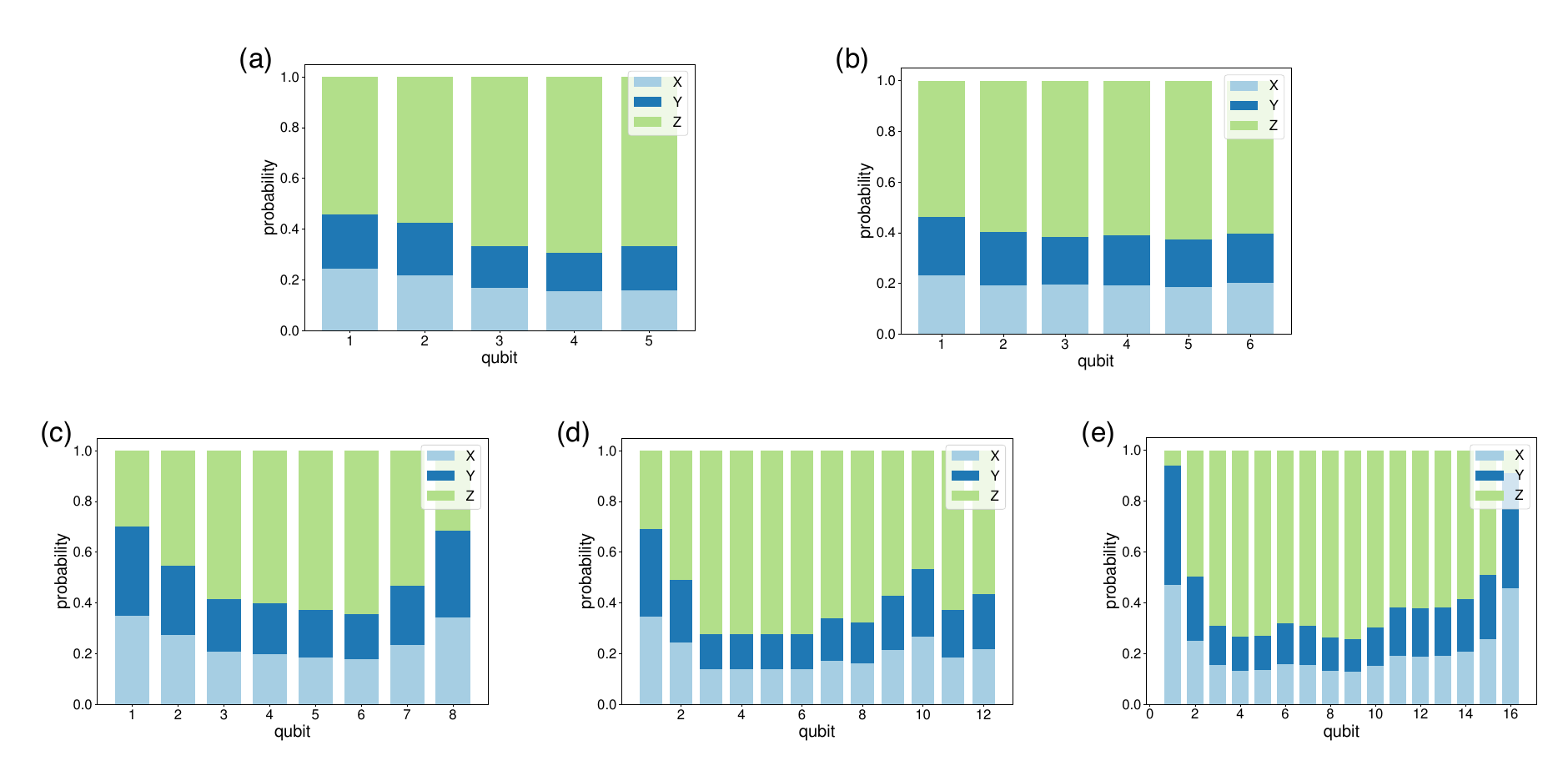}
    \caption{Query distributions in LBCS for Hamiltonians of different molecules. As the query distribution in LBCS is a product distribution, the marginal probability distribution over single-qubit Paulis are shown for each qubit. Molecules correspond to those in Table~\ref{table:molecules} and as shown are (a) \ce{H_2}(5 qubits, 3-21g, JW), (b) \ce{HeH^{+}} (6 qubits, 3-21g, JW), (c) \ce{HeH^{+}} (8 qubits, 6-31g, JW), (d) \ce{LiH} (12 qubits, sto6g, JW), and (e) \ce{N_2} (16 qubits, sto6g, JW).}
    \label{fig:LBCS_beta_molecules}
\end{figure}

We now visualize different instances of query distributions as obtained from the measurement methods of LBCS and decision diagrams. In Figure~\ref{fig:LBCS_beta_molecules}, we show the product query distributions obtained in LBCS by optimizing the variance of the energy estimate considering the state to be a maximally mixed state (Section~\ref{sec:meas_methods}).

Instead of showing illustrations of the decision diagrams (DD), we present relevant details of the optimized compact decision diagrams for different molecules in Table~\ref{tab:DD_beta_molecules}. It should be noted that the number of paths for each DD corresponds to the maximum number of unique measurement circuits that are being considered for each molecule. For example, the DD of \ce{LiH} only considers $810$ unique measurement circuits in contrast to CS for \ce{LiH} which would consider $3^{12} \approx 5.3 \times 10^5$ measurement circuits for high measurement budgets. Reducing the unique number of measurement circuits can help in reducing classical latencies as we mentioned in Section~\ref{sec:CHSCQCBench} and as observed in Section~\ref{sec:device_results}. As in the case of LBCS, the decision diagrams for each molecule are optimized by minimizing the variance of the energy estimate considering the state to be the maximally mixed state. We refer to resulting cost obtained from the optimized DD as the diagonal cost and is shown in Table~\ref{tab:DD_beta_molecules} with that for LBCS given as a reference. Finally, constructing and optimizing the decision diagram can be expensive. Hence, we report the classical computational runtimes required in obtaining the decision diagrams used in this work and in computing the classical pre-processing times as part of resource utilization (Tables~\ref{tab:resource_analysis_sim},\ref{tab:resource_analysis_dev}).

\begin{table}[ht!]
\small
\centering
{\renewcommand{\arraystretch}{1.2}
\begin{tabular}{|c|r|r|r|r|r|r|}
\hline
\textbf{Molecule} &
\multicolumn{1}{c|}{$\mid \text{\textbf{Nodes}} \mid$} &
\multicolumn{1}{c|}{$\mid \text{\textbf{Edges}} \mid$} &
\multicolumn{1}{c|}{$\mid \text{\textbf{Paths}} \mid$} &
\begin{tabular}[c]{@{}l@{}}\textbf{Diagonal cost} \\ \textbf{of Dec. Diag.}\end{tabular} &
\multicolumn{1}{c|}{\begin{tabular}[c]{@{}c@{}}\textbf{Diagonal cost of LBCS}\\ \textbf{(for reference)}\end{tabular}} &
\multicolumn{1}{c|}{\begin{tabular}[c]{@{}c@{}}\textbf{Computational runtime}\\ \textbf{of constructing DD [s]}\end{tabular}} \\ \hline
\begin{tabular}[c]{@{}c@{}}\ce{H_{2}}, 5 qubits \\ (3-21g, JW)\end{tabular}  & $40$ & $81$ & $43$ & $14.63$ & $26.59$ & $2.88 \times 10^1$ \\ \hline
\begin{tabular}[c]{@{}c@{}}\ce{HeH^{+}}, 6 qubits \\ (3-21g, JW)\end{tabular} &  $75$ & $155$  & $174$  & $17.11$ & $35.56$ & $5.45 \times 10^2$  \\ \hline
\begin{tabular}[c]{@{}c@{}}\ce{HeH^{+}}, 8 qubits \\ (6-31g, JW)\end{tabular} & $168$  & $337$  & $792$  & $16.69$ & $42.14$ & $3.54 \times 10^3$  \\ \hline
\begin{tabular}[c]{@{}c@{}}\ce{LiH}, 12 qubits \\ (sto6g, JW) \end{tabular} & $348$  & $600$  &  $810$ & $25.45$ & $42.44$ & $2.56 \times 10^4$  \\ \hline
\begin{tabular}[c]{@{}c@{}}\ce{N_2}, 16 qubits \\ (sto6g, JW)\end{tabular}  & $1118$  & $1551$  & $1137$  & $10836.12$ & $12121.62$ & $1.39 \times 10^5$  \\ \hline
\end{tabular}
}
\caption{Details of decision diagrams for different molecular Hamiltonians (Table~\ref{table:molecules}).  Number of paths in a decision diagram correspond to number of unique measurement circuits and contributes to classical latencies such as compilation time and circuit loading. Diagonal cost of a query distribution corresponds to the one-shot variance of the energy estimate considering $\rho$ to be the maximally mixed state. The reported computational runtimes account for both initialization and optimization of the decision diagrams.}
\label{tab:DD_beta_molecules}
\end{table}

\subsection{Convergence behavior of Adaptive Pauli Shadows}\label{app_sec:convergence_APS}
\begin{figure}[H]
    \centering
    \includegraphics[width=\textwidth]{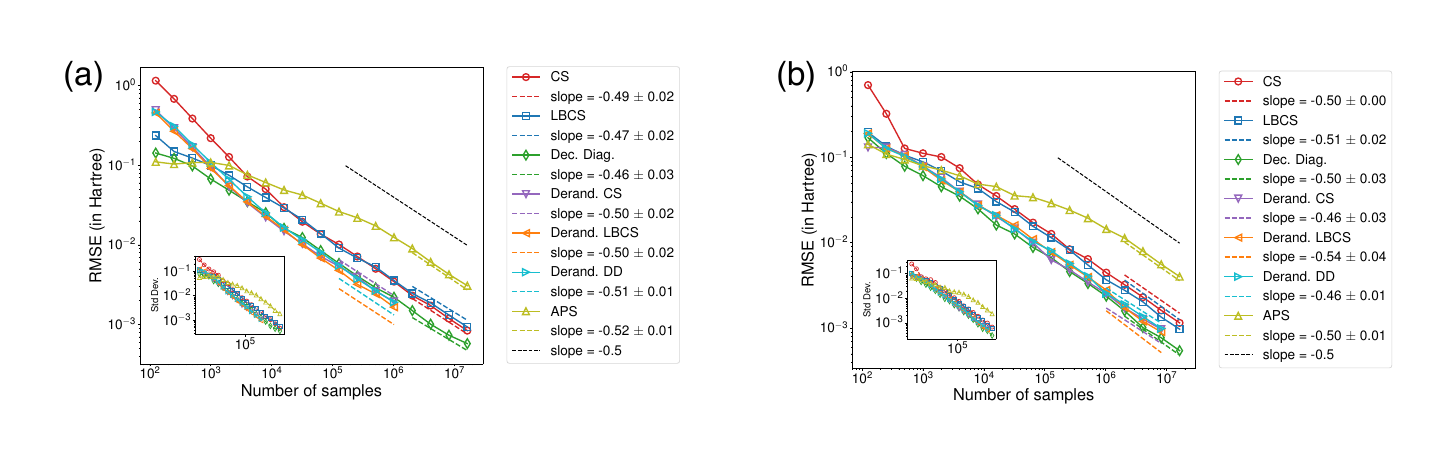}
    \caption{Comparison of RMSE achieved in numerical simulations by different measurement methods including APS in estimating $\Tr(\rho H)$ with $\rho$ set as the ground state and $H$ is the Hamiltonian of (a) tapered \ce{H_2} ($5$ qubits, 3-21g basis, JW encoding), and (b) tapered \ce{HeH^{+}} ($6$ qubits, 3-21g basis, JW encoding). RMSE is shown with the number of samples (or shots) made. The estimator for each measurement method is set to be the Bayesian estimator.}
    \label{fig:comparison_APS_meas_methods_H2_5q_HeH_6q}
\end{figure}

In Figure~\ref{fig:comparison_APS_meas_methods_H2_5q_HeH_6q}, we plot the trend of RMSE in estimating ground state energy of the tapered Hamiltonians from Table~\ref{table:molecules} in numerical simulation, considering for the measurement method of Adaptive Pauli Shadows (APS) and other measurement methods. It has been shown earlier in \cite{hadfield2021adaptive} that APS outperforms other measurement methods at the low measurement budget of $10^3$ shots and this is observed here as well. However, this does not continue for higher measurement budgets and we instead observe a weird convergence behavior for APS. Such behavior has been reported earlier in \cite{shlosberg2023adaptiveestimation}. Hence, APS was not included as a candidate measurement method in CSHOREBench.

\end{document}